\documentclass[journal]{IEEEtran}
\usepackage{epsfig,graphics,graphicx,subfigure,amssymb,amstext,amsmath,algorithm,
algorithmic,wrapfig,multirow}
\usepackage{array}
\usepackage{cite}
\usepackage{url}
\usepackage{times}
\usepackage{lipsum}
\usepackage{float}
\usepackage{placeins}
\usepackage{textcomp}
\usepackage[font=small,bf]{caption}
\usepackage{hyperref}
\usepackage{flushend}
\usepackage{xcolor}
\usepackage{tabularx}
\usepackage{bm}
\usepackage{enumerate}
\usepackage{tikz}
\usepackage{pgfplots}
\usepackage{enumerate}
\usepackage{epstopdf}
\usetikzlibrary{shapes,arrows}
\usepackage[utf8]{inputenc}

\def\beq{\begin{equation}}
\def \eeq{\end{equation}}
\def\beqa{\begin{eqnarray}}
\def\eeqa{\end{eqnarray}}
\def\beqan{\begin{eqnarray*}}
\def\eeqan{\end{eqnarray*}}

\def\C{{\mathbb{C}}}
\def\argmin{\mathop{\mathrm{arg\,min}}}

\def\x{\times}

\def\SNR{\mbox{\small \sffamily SNR}}

\def\tm1{t\! - \! 1}
\def\tp1{t\! + \! 1}

\def\rbf{\mathbf{r}}

\def\sbf{\mathbf{s}}

\def\ubf{\mathbf{u}}

\def\vbf{\mathbf{v}}

\def\wbf{\mathbf{w}}

\def\Hbf{\mathbf{H}}
\def\Hhat{\widehat{H}}
\def\Ibf{\mathbf{I}}

\def\ellhat{\widehat{\ell}}

\def\psibf{{\boldsymbol \psi}}

\def\taubf{{\boldsymbol \tau}}

\newif\ifconf
\conffalse

\newif\ifthreebitq
\threebitqtrue

\newif\ifonecol
\onecolfalse

\IEEEoverridecommandlockouts

\renewcommand{\footnoterule}{%
  \kern -3pt
  \hrule width \columnwidth height 0.5pt
  \kern 3pt
}
\begin{document}

\title{Initial Access in Millimeter Wave\\ Cellular Systems}

  \author{
        C. Nicolas Barati~\IEEEmembership{Student Member,~IEEE},
        S. Amir Hosseini~\IEEEmembership{Student Member,~IEEE},
        Marco Mezzavilla~\IEEEmembership{Member, IEEE},
        Thanasis Korakis,~\IEEEmembership{Senior Member,~IEEE},
        Shivendra S. Panwar,~\IEEEmembership{Fellow,~IEEE},\\
        Sundeep Rangan,~\IEEEmembership{Fellow,~IEEE},
        Michele Zorzi,~\IEEEmembership{Fellow, ~IEEE},
        \thanks{This material is based upon work supported by the National Science
        Foundation under Grants No.~ 1116589, 1237821 and 1547332 as well as generous support
        from NYU industrial affiliates.}
        \thanks{
            C. N. Barati (email: nicolas.barati@nyu.edu),
            S. Amir Hosseini (email: amirhs.hosseini@nyu.edu),
            Marco Mezzavilla (email: mezzavilla@nyu.edu),
            S. Rangan (email: srangan@nyu.edu),
            Thanasis Korakis (email: korakis@nyu.edu),
            S. S. Panwar (email: panwar@nyu.edu),
            are with the NYU WIRELESS, Tandon School of Engineering,
            New York University, Brooklyn, NY.
            M. Zorzi (email: zorzi@dei.unipd.it) is with the Department of Information Engineering,
            University of Padova, Italy.}
         \thanks{ A preliminary version of this paper was presented at Asilomar 2015 \cite{BarHosAsilomar:15}.
         }
    }

\maketitle

\begin{abstract}
The millimeter wave (mmWave) bands have recently attracted considerable
interest for next-generation cellular
systems due to the massive spectrum at these frequencies.
However, a key challenge in designing mmWave cellular systems is 
initial access -- the procedure by which a mobile device establishes an initial link-layer connection to a base station cell.
MmWave communication relies on highly directional transmissions and the initial access procedure must thus provide 
a mechanism by which initial transmission directions can be searched in a potentially large angular space. 
Design options are compared considering different scanning and signaling procedures to evaluate access delay and system overhead. 
The channel structure and multiple access issues are also considered.
The results of our analysis demonstrate significant benefits of low-resolution fully digital architectures 
in comparison to single stream analog beamforming.
\end{abstract}


\section{Introduction}

The millimeter wave (mmWave) bands -- roughly corresponding to
frequencies above 10~GHz --  are a new frontier for
cellular wireless communications~\cite{KhanPi:11-CommMag,rappaportmillimeter,RanRapE:14,andrews2014will,ghosh2014millimeter,DehosG:14}.
These frequency bands offer orders of magnitude more spectrum than the
congested bands in conventional UHF and microwave frequencies below 3~GHz.
In addition, advances in CMOS RF circuits combined with
the small wavelengths of mmWave frequencies enable large numbers of
electrically steerable antenna
elements to be placed in a picocellular access point or mobile.
 These high-dimensional antenna arrays can provide further gains
via adaptive beamforming and spatial multiplexing.
Preliminary capacity estimates demonstrate that
this combination of massive bandwidth with large numbers of
spatial degrees of freedom can enable orders of magnitude
increases in capacity over current cellular systems
\cite{AkdenizCapacity:14,BaiHeath:15}.

However, much remains to be designed to enable
cellular systems to achieve the potential of the mmWave bands \cite{shokri2015millimeter}.
One issue often neglected in the discussions on future cellular networks using mmWave is that of \emph{control plane latency.} 
That is, the delay a mobile device (or user equipment (UE) in 3GPP terminology) experiences during the transition from idle mode to connected state.
This transition may occur much more often than in current LTE deployments for two reasons: i) \emph{Intermittency of the links:} mmWave links are acutely susceptible to shadowing.
Furthermore, mmWave cells are projected to be small in size and the coverage may be ``spotty''. In this case, the coverage holes will be filled by a 4G macro-cell.
Therefore, frequent inter-mmWave and cross-technology
handovers are expected to occur.
ii) \emph{Frequent idle mode cycles:} Operating at extremely high
frequencies and with wide bandwidths may quickly drain the UE's battery. Therefore,
a more aggressive use of idle mode may be necessary.
 
This paper considers the basic problem of \emph{initial access (IA)}
-- the procedure by which a UE
discovers a potential mmWave cell
and establishes a link-layer connection \cite{giordani:16multi}.
It is this very procedure that will be triggered in case of an intermittent
link or recovery from idle mode.
Hence, addressing this problem will have significant impact on the control plane latency.

Initial access is a basic prerequisite to any communication
and is an essential component of all cellular systems.  However,
mmWave communication relies heavily on highly-directional transmissions to overcome
the large isotropic pathloss and the
use of directional transmissions significantly complicates initial access.
In addition to detecting the presence of the base station and access request from the
UE, the mmWave initial access procedure
must provide a mechanism by which both the UE and the base station (BS) can determine suitable beamforming (BF)
directions on which subsequent directional communication can be carried out.
With very high BF gains, this angular search can significantly
slow down the initial access, due to the potentially large beam search space.
This increase in delay goes against one of the main
objectives of mmWave systems, which is to dramatically reduce both data plane
and control plane latency~\cite{BocHLMP:14,levanen2014radio}.

\subsection{Contributions} \label{sec:contrib}
This paper presents several different design options for mmWave initial access.
We consider various procedures, using
the basic steps used in the LTE standard as a reference
but with this major modification: beside detecting each other,
\emph{both the UE and the BS learn the initial BF
directions.}
The design options consider different methods for transmitting
and receiving the synchronization (Sync) signals from the BS and the random access (RA) request (also called preamble) from the UE.

In our previous work \cite{BarHosCellSearch:TWC15}, we explored the problem of synchronization (downlink) in a mmWave cell using random BF TX/RX.
We were interested in the boundaries of the SNR region where the Sync signal is detectable.
This work is a major extension to \cite{BarHosCellSearch:TWC15} as it looks at the whole IA procedure (through five basic design options) which includes both Sync and UE discovery by the BS.
We derive an optimal detector based on the assumption that
each side (BS /UE) has a fixed set of BF directions in which they
can transmit and receive signals.   
Rather than just the SNR regimes, here we are interested in the overall delay of each design option and we evaluate this delay as a function of the system overhead, thereby determining the control plane latency.

Due to the wide bandwidths and large number of antenna elements in the mmWave range, it may not be possible from a power consumption perspective
for the UE to obtain high rate digital samples
from all antenna elements~\cite{KhanPi:11,Abbas:16BF}.
Most proposed designs perform BF in analog (at either RF or IF)
prior to the analog to digital (A/D) conversion
\cite{KohReb:07,KohReb:09,GuanHaHa:04,Heath:partialBF,SunRap:cm14}. 
A key limitation of these
architectures is that they permit the
UE to ``look" in only one or a small number of directions at a time.
In this work, in addition to analog BF we consider
a theoretical fully digital architecture that can look in all directions at once.
To compensate for the high power consumption of this fully digital architecture,
we consider a low-resolution design where each antenna is quantized at very
low bit rates (say 2 to 3 bits per antenna) as proposed by
\cite{Madhow:ADC,Madhow:largeArray} and related methods in \cite{hassan2010analog}.

It is worth noting that throughout this work we assume a standalone mmWave system.
Some recent papers such as \cite{Nitsche:15} discuss assisted peer discovery, proposing out of band (sub $6$~GHz) detection by using overheard packets.
Others use an ``external localization service''\cite{capone:15context,jung:15cellDtct,Abbas:16context} to get positioning information, provided perhaps using GPS.
Unfortunately, in most of these works 
discovery 
requires
line of sight (LOS) paths while we assume that establishing NLOS links is also achievable.
Therefore, in our simulation we evaluate the proposed design option via a realistic LOS/NLOS channel model. 

The rest of the paper is organized as follows.
In Section~\ref{desopt} we describe initial access 
in a mmWave cellular system while drawing clear parallels with the current LTE procedure. 
We provide various design options for the two commencing phases 
of initial access, namely, synchronization and random access.
In Section~\ref{sec:sigdetect} we present our assumed channel model and theoretically 
derive an optimal detector for the analog BF case,
arguing that the same detector can be used for the digital beamforming case as well.
In Section~\ref{sec:evalan} we evaluate through simulation the design options presented earlier and 
finally in Section~\ref{concl} we summarize our major findings.  

\section{Design Options for mmWave Initial Access} \label{desopt}

\subsection{Initial Access Procedure Steps}

\ifonecol

\begin{figure}[!tbp]
  \centering
  \begin{minipage}[t]{0.45\textwidth}
    \includegraphics[width=\textwidth]{IAtx.pdf}
    \caption{Directional initial access procedure.}
    \label{fig:initAccess}
  \end{minipage}
  \hfill
  \begin{minipage}[t]{0.54\textwidth}
    \includegraphics[width=\textwidth]{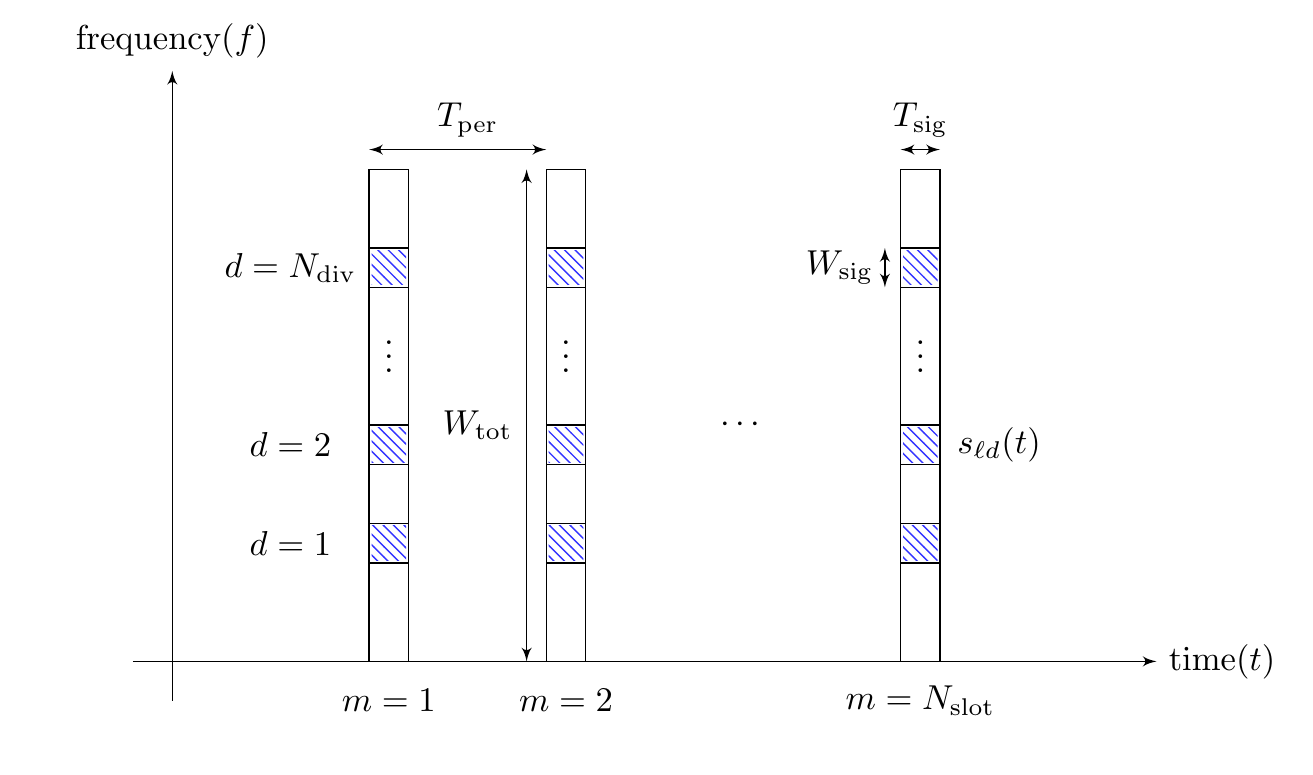}
    \caption{ Periodic transmission of narrowband synchronization signals from the BS.}
    \label{fig:acsSig}
  \end{minipage}

\end{figure}

\else
\begin{figure}
\begin{center}
\pgfmathsetmacro{\xbs}{0}
\pgfmathsetmacro{\xue}{5}
\pgfmathsetmacro{\dy}{0.5}  
\pgfmathsetmacro{\ysync}{9}  
\pgfmathsetmacro{\ypre}{6}  
\pgfmathsetmacro{\yacs}{3}  
\pgfmathsetmacro{\yreq}{1}  
\pgfmathsetmacro{\ydat}{-1}  
\begin{tikzpicture}[scale=0.4]
    \draw[thick, -] (\xbs,-3.5) -- (\xbs,10) node [above] {BS};
    \draw[thick, -] (\xue,-3.5) -- (\xue,10) node [above] {UE};

    \draw[->] (\xbs,\ysync) -- (\xue,\ysync-\dy)
        node[draw=none,fill=none,midway,below,yshift=0.25cm] {$\vdots$}
        node[draw=none,fill=none,midway,above,yshift=0] {\textcircled{\tiny 0}} ;
    \draw[->] (\xbs,\ysync-1.25) -- (\xue,\ysync-\dy-1.25);
    \node at (\xbs,\ysync) [left,text width=2cm,font=\scriptsize,align=right]
        {Send periodic sync signal};
    \node at (\xue,\ysync-1) [right, text width=2cm,font=\scriptsize] {Detect BS \& find UE BF direction};

    \draw[->] (\xue,\ypre) -- (\xbs,\ypre-\dy)
        node[draw=none,fill=none,midway,above,yshift=0] {\textcircled{\tiny 1}} ;
    \node at (\xue,\ypre) [right, text width=2cm,font=\scriptsize,yshift=-0.25cm]
        {Send randomly selected RA preamble};
    \node at (\xbs,\ypre-\dy) [left, text width=2cm,font=\scriptsize,align=right]
        {Detect RA preamble \& find BS BF direction};

    \draw[->] (\xbs,\yacs) -- (\xue,\yacs-\dy)
        node[draw=none,fill=none,midway,above,yshift=0] {\textcircled{\tiny 2}} ;
    \node at (\xbs,\yacs) [left, text width=2cm,font=\scriptsize,align=right] {
        Send RAR with RNTI, UL grant \& other info};

    \draw[->] (\xue,\yreq) -- (\xbs,\yreq-\dy)
        node[draw=none,fill=none,midway,above,yshift=0] {\textcircled{\tiny 3}} ;
    \node at (\xue,\yreq) [right, text width=2cm,font=\scriptsize]
        {Send connection request};

    \draw[implies-implies,double equal sign distance] (\xue,\ydat) -- (\xbs,\ydat)
        node[draw=none,fill=none,midway,above,yshift=0] {\textcircled{\tiny 4}}        
        node[draw=none,fill=none,midway,below,text width=2cm,align=center,font=\scriptsize] {Scheduled communication};

\end{tikzpicture}
\end{center}
\caption{Directional initial access procedure.} \label{fig:initAccess}
\end{figure}
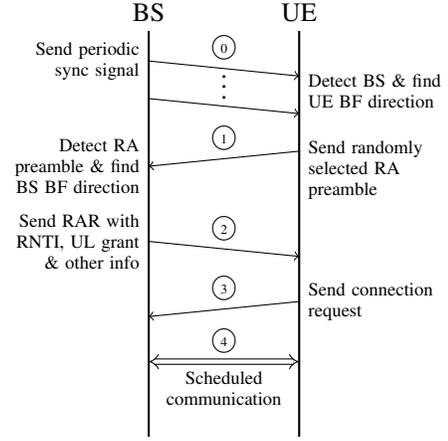

\begin{figure}
  \includegraphics[width=0.54\textwidth]{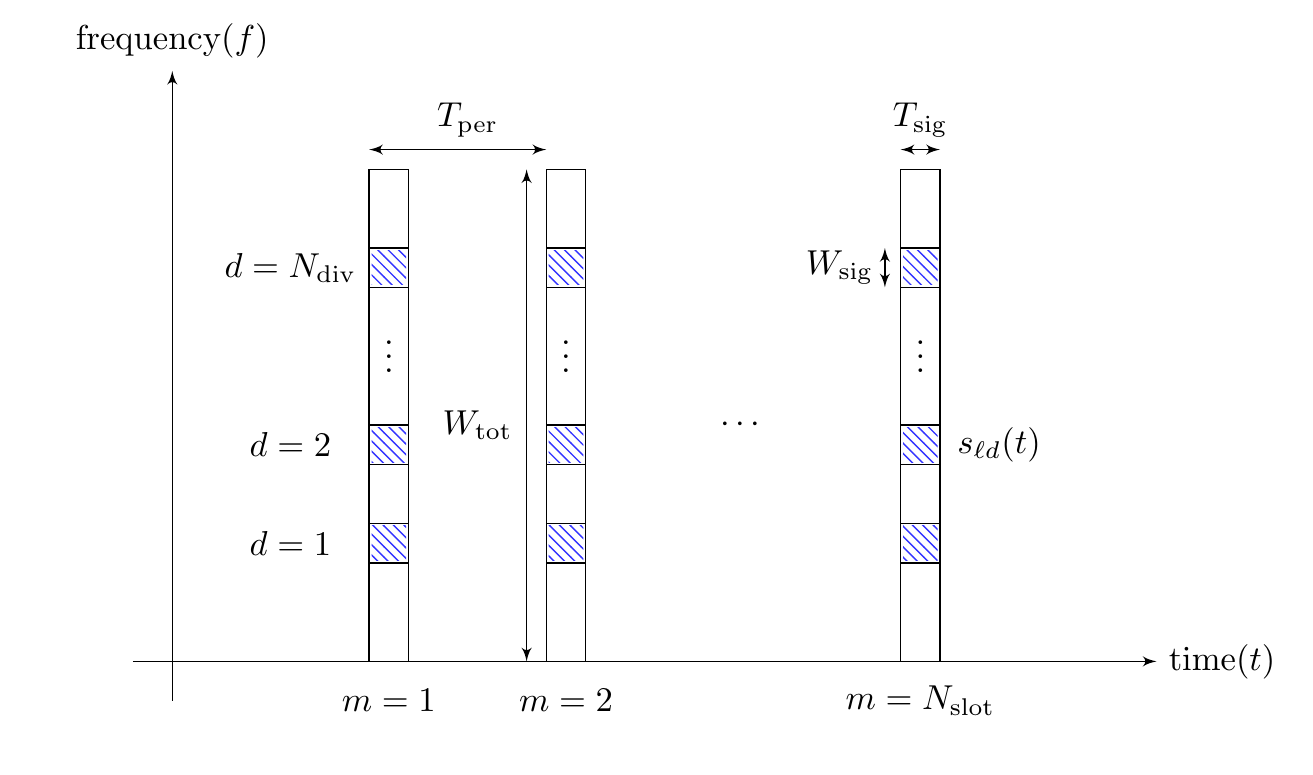}
    \caption{ Periodic transmission of narrowband synchronization signals from the BS.}
    \label{fig:acsSig}
  \end{figure}

\fi

All the design options considered for initial access follow the same basic steps
as shown in Figure~\ref{fig:initAccess}.
At a high level, these steps are identical to the methods used in 3GPP LTE, which are described in the specifications
\cite{3GPP36.300,3GPP36.321,3GPP36.331} as well as
any standard text such as \cite{Dahlman:07}.
However, the depicted procedure needs major modifications for mmWave
to enable both the UE and the BS to determine the initial BF directions in addition to detecting the presence of the BS and the access request from the UE. 
The steps are as follows.
Note that since the most challenging task of mmWave IA is to determine 
the spatial signatures of the BS and the UE, 
we focus on the first two steps and do not elaborate on the rest.

\begin{enumerate}[1)]
\item[0)] \emph{Synchronization signal detection:}
Each cell periodically transmits synchronization signals that the UE can scan
to detect the presence of the base station and obtain the downlink frame timing.
In LTE, the first synchronization signal to detect is the
Primary Synchronization Signal (PSS).  For mmWave, the synchronization signal will also
be used to determine the UE's BF direction, which is related to the angles of arrival
of the signal paths from the BS.
Critical to our analysis, we will assume that the UE only attempts to learn the long-term
BF directions \cite{Lozano:07}, which depend
only on the macro-level scattering paths and do not vary with small scale fading. 
The alternative, i.e., instantaneous beamforming, would require channel state information (CSI)
at both TX and RX which may not be feasible at the initial stages of random access and also due to high Doppler frequencies, a point made in \cite{AkdenizCapacity:14}. 
As a result, the long-term BF directions will be stable over much longer periods
and thus will be assumed to be constant over the duration of the initial access
procedure.


\item \emph{RA preamble transmission:} Similar to LTE, we assume that the uplink contains
dedicated slots exclusively for the purpose of random access (RA) messages.
After detecting the synchronization signals and decoding the broadcast
messages, the location of these RA slots is known to the UE.
The UE randomly selects one of a small number (in LTE, there are up to 64)
of waveforms, called \emph{RA preambles}, and transmits the preamble in
one of the RA slots.  In all design options we consider below, the
UE BF direction is known after step 0, so the RA preamble can be transmitted directionally,
thereby obtaining the BF gain on the UE side.
The BS will scan for the presence of the RA preamble and will also learn
the BF direction at the BS side.
As we discuss below, the method by which
the BS will learn the BF direction will depend on the RA procedure.

\item \emph{Random access response (RAR):}
Upon detecting a RA preamble, the base station transmits
a random access response to the UE indicating the index
of the detected preamble.  At this point, both the BS and the UE know the
BF directions so all transmissions can obtain the full BF gain.
The UE receiving the RAR knows its
preamble was detected.

\item \emph{Connection request:} After receiving RAR, the UE desiring initial access
will send some sort of connection request message
(akin to ``Radio Resource Control (RRC) connection request" in LTE)
on the resources scheduled in the uplink (UL) grant in the RAR.

\item \emph{Scheduled communication:}
At this point, all subsequent communication can occur on scheduled channels
with the full BF gain on both sides.
\end{enumerate}

\begin{table*}
\begin{center}
\begin{tabular}{|>{\raggedright}p{1 cm }|>{\raggedright}p{2 cm }|>{\raggedright}p{2 cm}|>{\raggedright}p{6 cm}|}
  \hline
  \textbf{Step} & \textbf{Item} & \textbf{Option} & \textbf{Explanation}
  \tabularnewline \hline
\multirow{2}{*} {0} &
\multirow{2}{*} {Sync BS TX} & Omni & BS transmits the downlink synchronization signal in
a fixed, wide angle beam pattern to cover the entire cell area.   \tabularnewline \cline{3-4}

    & & Directional & BS transmits the downlink synchronization signal in
    time-varying narrow-beam directions to sequentially scan the angular space.  \tabularnewline \hline

\multirow{2}{*} {0} &
\multirow{2}{*} {Sync UE RX}
    & Directional & UE listens for the downlink synchronization signal
    in time-varying narrow-beam directions with analog BF that sequentially scans
    the angular space. \tabularnewline \cline{3-4}

  &   & Digital RX & UE has a fully digital RX and can thereby receive in all
    directions at once.  \tabularnewline \hline

\multirow{2}{*} {1} &
\multirow{2}{*} {RA BS RX}
    & Directional & In each random access slot,
    the BS listens for the uplink random access signal
    in time-varying narrow-beam directions using analog BF.  The directions
    sequentially scan the angular space. \tabularnewline \cline{3-4}

  &   & Digital RX & The BS has a fully digital RX and can thereby receive in all
    directions in the random access slot at once.  \tabularnewline \hline
\end{tabular}
\caption{Design options for each stage}
\label{tab:RachOpSteps}

\end{center}
\end{table*}

\subsection{Synchronization and Random Access Signals}

We assume that synchronization and random access signals are transmitted in relatively narrowband waveforms
in periodically occurring intervals as shown in Figure~ \ref{fig:acsSig}.
As shown in the figure, the synchronization signal is transmitted periodically
once every $T_{\rm per,sync}$ seconds for a duration of $T_{\rm sig,sync}$
in each transmission.
In LTE, the primary synchronization signal (PSS) is transmitted once every
5~ms for a duration of one OFDM symbol ($71.3~\mu$s).
In this work, we will consider potentially different periods and signal lengths.
We assume that the random access slots
are located once every $T_{\rm per,RA}$ seconds and that the random access signals
are of length $T_{\rm sig,RA}$.  In LTE, the frequency of the
random access slots is configurable and they are located at least once every
10~ms.  Both the synchronization and the random access signals are assumed to be
relatively narrowband with bandwidths $W_{\rm sig,sync}$ and $W_{\rm sig,RA}$, respectively. 
Note that in general, the periodicity, signal length, and signal bandwidth can be different for the
synchronization and random access phases, resulting in different delays and overheads in the two cases.
In this paper however, for the sake of simplicity and also to make the comparison between different design options more tractable, 
we take them to be equal for both phases and will refer to them simply as $T_{\rm per}$, $T_{\rm sig}$, and $W_{\rm sig}$, respectively.


\ifonecol
\begin{table}
\begin{center}
\begin{tabular}{|p{1cm}|p{1.55cm}|p{1.6cm}|p{1.4cm}|c|c|p{3cm}|p{3.8cm}|}
  \hline
  \multirow{2}{3em}{\bf Option} & \multirow{2}{5.6em}{\bf Sync BS TX} & \multirow{2}{5.6em}{\bf Sync UE RX} & \multirow{2}{5.4em}{\bf RA BS RX}
    & \multicolumn{2}{|c|}{$\bf{L}$} & \multirow{2}{10em}{\bf How the UE learns the BF direction} & \multirow{2}{10em}{\bf How the BS learns the BF direction} \\
    &&&& Sync & RA &&\\
  \hline

(a) DDO & Directional & Directional & Omni & $N_{\rm tx}N_{\rm rx}$& 1& Directional scanning & The time slot index
on which the UE received the sync signal is encoded
in the RA preamble index (which may reduce the number of available preamble waveforms).
Thus, the BS knows the TX direction in which the sync signal was received.\\
\hline

(b) DDD & Directional & Directional & Directional & $N_{\rm tx}N_{\rm rx}$ & $N_{\rm tx}$ & Same as (a) & The BS scans the directions for the RA preamble, the BF direction can be learned from
the direction in which the RA preamble is received.\\
\hline

(c) ODD & Omni & Directional & Directional & $N_{\rm rx}$ & $N_{\rm tx}$ & Same as (a) & Same as option (b).\\
\hline

(d) ODDig & Omni & Directional & Digital RX  & $N_{\rm rx}$ & $N_{\rm tx}$ & Same as (a)
& With digital BF at the BS, the direction can be learned
from the spatial signature of the received preamble.\\
\hline

(e) ODigDig & Omni & Digital RX &  Digital RX & 1 & 1 & With digital BF at the UE, the
direction can be learned from the spatial signature of the
received sync signal & Same as option (d).\\
\hline

\end{tabular}
\caption{Design option combinations for initial access. The fifth and sixth columns are the size of the angular domain $L$ that has to be scanned for the Sync and the RA preamble detection respectively.}
\label{tab:RachOpCombo}
\end{center}
\end{table}

\else

\begin{table*}
\begin{center}
\begin{tabular}{|l|c|c|c|c|c|p{3 cm}|p{3.8 cm}|}
  \hline
  \multirow{2}{3em}{\bf Option} & \multirow{2}{5.4em}{\bf Sync BS TX} & \multirow{2}{5.6em}{\bf Sync UE RX} & \multirow{2}{5.4em}{\bf RA BS RX}
    & \multicolumn{2}{|c|}{$\bf{L}$} & \multirow{2}{10em}{\bf How the UE learns the BF direction} & \multirow{2}{9em}{\bf How the BS learns the BF direction} \\
    &&&& Sync & RA &&\\
  \hline

(a) DDO & Directional & Directional & Omni & $N_{\rm tx}N_{\rm rx}$ & 1& Directional scanning & The time slot index
on which the UE received the sync signal is encoded
in the RA preamble index (which reduces the number of available preamble waveforms).
This way, the BS knows the TX direction in which the sync signal was received.
Alternatively, the index of the time slot of the sync signal can be encoded implicitly
in the RA preamble slot, which increases the access delay since the UE must wait for
the appropriate access slot, but keeps all RA preamble waveforms available.\\
\hline

(b) DDD & Directional & Directional & Directional & $N_{\rm tx}N_{\rm rx}$ & $N_{\rm tx}$ & Same as (a) & Since the BS scans the directions for the RA preamble, the BF direction can be learned from
the direction in which the RA preamble is received.\\
\hline

(c) ODD & Omni & Directional & Directional & $N_{\rm rx}$& $N_{\rm tx}$ & Same as (a) & Same as option (b).\\
\hline

(d) ODDig & Omni & Directional & Digital RX  & $N_{\rm rx}$ & 1 & Same as (a)
& With digital BF at the BS, the direction can be learned
from the spatial signature of the received preamble.\\
\hline

(e) ODigDig & Omni & Digital RX &  Digital RX & 1 & 1 & With digital BF at the UE, the
direction can be learned from the spatial signature of the
received sync signal & Same as option (d).\\
\hline

\end{tabular}
\caption{Design option combinations for initial access. The fifth and sixth columns are the size of the angular domain $L$ that has to be scanned for the Sync and the RA preamble detection respectively.}
\label{tab:RachOpCombo}
\end{center}
\end{table*}

\fi

\subsection{Learning the BF Directions}\label{sec:learnBF}

While the basic procedure steps for initial access are similar to that used currently in LTE,
the mmWave initial access procedure must include a method by which the
base station and the UE can learn the directions of communication.
The key modifications to enable this learning would
occur in steps 0 and 1 of the procedure in Figure~\ref{fig:initAccess}.
In step 0, the UE must learn its BF direction, and in step 1 the BS learns the BF direction on its side.
As stated earlier, these steps are the most challenging ones, hence we will focus on them.
Table~\ref{tab:RachOpSteps} shows some options for three items in this procedure,
namely: (i) the manner in which the synchronization signal in step 0 is transmitted
by the BS;
(ii) the manner in which the synchronization signal is received at the UE; and
(iii) the manner in which the random access preamble from the UE
is received at the BS in step 1.
For each item, we consider two possible options.  For example,
for the manner in which the synchronization signal is transmitted,
we consider a fixed omni-directional transmission or a sequential scanning with directional
transmissions.
Note that since it is assumed that by the end of step 0 the UE has learned the correct direction of arrival from the BS , in step 1 it will always beamform in that direction to transmit the random access preamble.

Now, since there are two options for each of the three items in Table~\ref{tab:RachOpSteps},
there are a total of eight design option combinations.
However, for reasons that will become clear in the next section, we only consider five of these options.
We will use the following nomenclature for these five options:
They are named after the manner the Sync signal is transmitted (omni or analog directional) and how both the Sync and RA signals are received (analog directional or digital). 
For example, a design where the Sync signal is transmitted and received in analog directional while the RA preamble is received in the same manner is called \emph{DDD}; and one where the signal is transmitted omni-directionally and received with digital BF in the Sync phase, and the RA preamble is received with digital BF is called \emph{ODigDig}.
We summarize the functionality and characteristics of these option combinations in Table~\ref{tab:RachOpCombo}.
This table includes the designs proposed in the recent works
\cite{jeong2015random,shokri2015millimeter}.

\subsection{Sequential Beamspace Scanning}\label{sec:seqBeamScan}
To analyze the different design options, we first need to quantify the
time it takes to detect the synchronization and random access signals
as a function of the  SNR and signal overhead.
We can analyze both the synchronization and RA phase with the same analysis.
Suppose a transmitter (TX) repeatedly broadcasts some known signal once every $T_{\rm per}$
seconds as illustrated in Figure~\ref{fig:acsSig}.
For frequency diversity,
we assume that each transmission consists of $N_{\rm div}$ \emph{subsignals}
transmitted in different frequency locations 
and that subsignals and their frequency
locations are known to the receiver.
The subsignals are assumed to be mostly constrained to some
small time-frequency region of size $T_{\rm sig} \times W_{\rm sig}$.
The goal of the receiver is
to detect the presence of the signal, and, if it is present, detect
its time and angle of arrival.


This model can be applied to analyze either the synchronization or the random
access phase.
In the synchronization phase (Step 0 of Figure~\ref{fig:initAccess}),
the TX will be the base station and the signal is the synchronization signal.
In the random access phase (Step 1 of Figure~\ref{fig:initAccess}),
the TX will be the UE and the signal will be its random access preamble.

Now, for the moment, suppose that both the TX and the RX
can perform only analog BF, so that they can only align their arrays
in one direction at a time.
We will address searching with hybrid and digital BF later
-- see Section~\ref{sec:hybDig}.
To detect the synchronization signal,
we assume that the TX and the RX cycle through $L$ possible TX-RX BF
direction pairs.
We will call each such cycle of $L$ transmissions a ``directional scan" or `` scan cycle''. 
Since the transmission period is $T_{\rm per}$ seconds, each scan cycle will take $LT_{\rm per}$ seconds.
We index the transmissions in each scan cycle by $\ell=1,\ldots,L$, and let $\ubf_\ell$
and $\vbf_\ell$ be the RX and TX beamforming vectors applied during the
$\ell$-th transmission in each scan cycle. The same beamforming vectors are applied
to all subsignals in each transmission.

There are a large number of options for selecting the $L$ direction pairs
$(\ubf_\ell,\vbf_\ell)$ to use
in each scan cycle -- see, for example, \cite{SunRap:cm14} for a summary of some
common methods.  In this work,
we will assume a simple \emph{beamspace} sequential search.
Given a linear or 2D planar array with $N$ antennas, the spatial signature of any plane
wave on that array is given by the superposition of $N$ orthogonal directions called
the beamspace directions.  Each beamspace direction corresponds to a direction of
arrival at a particular angle.  In beamspace scanning, the TX and RX directions
that need to be searched are selected from the orthogonal beamspace directions.
If the TX and RX have $N_{\rm tx}$ and $N_{\rm rx}$ antennas, respectively, then
the number $L$ of directions to scan is given by:
\begin{itemize}
\item $L=N_{\rm tx}N_{\rm rx}$ if both the TX and RX scan over all the beamspace directions;
\item $L=N_{\rm tx}$ if only the TX scans while the RX uses an omni-directional or fixed
antenna pattern;
\item $L=N_{\rm rx}$ if only the RX scans while the TX uses an omni-directional or fixed
antenna pattern;
\item $L=1$ if both the TX and RX use omni-directional (or fixed antenna patterns) or digital BF. In the latter case, unlike in analog BF the digital receiver can ``look'' into all $N_{\rm rx}$ directions simultaneously while maintaining the RX directivity gain.    
\end{itemize}
The TX and RX continue repeating the transmission scans until either
the signal is detected or the procedure times out.
Therefore, the size of the angular space $L$ is directly connected to
the delay of signal detection since more time is needed to cover all the angular pairs as $L$ grows.
In Table~\ref{tab:RachOpCombo}, for each design we have put the respective number of directions $L$ in both Sync and RA.
Note that RA's $L$ is always less than or equal to Sync's $L$ since after step 0, 
the position of the BS is assumed to be known and only one side may need to scan the angular space.



In Section~\ref{sec:learnBF} we mentioned that we present only five designs out of the eight possible combinations of Table~\ref{tab:RachOpSteps}. 
The explanation is that every one of the three designs omitted share the same number of directions to scan $L$ with some of the five basic designs and their performance in delay was assessed to be no different than the ones presented in Table~\ref{tab:RachOpCombo}.

\section{Signal Detection under a Sequential Beamspace Scanning} \label{sec:sigdetect}

\subsection{Generalized likelihood ratio (GLRT) Detection}\label{sec:glrtDtct}

Given the above transmissions, the RX must determine whether
the signal is present, and, if it is present, determine the direction
of maximum energy.  To make this decision, suppose that the RX
listens to $K$ scan cycles for a total of $LK$ transmissions.
Index the transmissions within each scan cycle by $\ell=1,\ldots,L$ and the
cycles themselves by $k=1,\ldots,K$.
We assume that each transmission occurs in some signal space with $M$
orthogonal degrees of freedom
and we let $\sbf_{\ell d} \in \C^M$ be the signal space representation
of the $d$-th subsignal transmitted in the $\ell$-th transmission
in each scan cycle where $d=1,\ldots,N_{\rm div}$.
If the subsignal can be localized to a time interval
of length $T_{\rm sig}$ and bandwidth $W_{\rm sig}$, then the number
of degrees of freedom is approximately $M \approx T_{\rm sig}W_{\rm sig}$.

We assume that in the $\ell$-th transmission within each scan cycle,
the RX and TX
apply the beamforming vectors $\ubf_\ell \in \C^{N_{\rm rx}}$ and $\vbf_\ell \in \C^{N_{\rm tx}}$.
We assume that the received complex baseband signal in the $d$-th subsignal
in the $\ell$-th transmission and $k$-th scan cycle can be modeled as
\beq \label{eq:ruv}
    \rbf_{k\ell d} = \ubf^*_\ell\Hbf_{k\ell d}\vbf_\ell \sbf_{\ell d} + \wbf_{k\ell d},
    \quad
    \wbf_{k\ell d} \sim {\mathcal N}(0,\tau_{k\ell d}\Ibf_M),
\eeq
where $\rbf_{k\ell d} \in \C^M$ is the signal space representation of the
received subsignal after beamforming;
$\Hbf_{k\ell d}$ is the complex MIMO fading channel matrix during the transmission, and
$\wbf_{k\ell d}$ is complex white Gaussian noise (WGN)
with some variance $\tau_{k\ell d}$ in each dimension.
We have assumed that the signal in each transmission is
transmitted in a sufficiently small time and frequency bandwidth so
that the channel can be approximated as constant and flat within that transmission.

Our first key simplifying assumption is that the true channel between the TX and RX,
if it exists, is described by a single LOS path exactly aligned with
one of the $L$ TX-RX beamspace directions.  That is,
\beq \label{eq:HLOS}
    \Hbf_{k\ell d} = h_{k\ell d}\ubf_{\ell_0}\vbf_{\ell_0}^*,
\eeq
so that the channel's TX direction $\vbf_{\ell_0}$ and RX direction $\ubf_{\ell_0}$
are each aligned exactly along one of the TX and RX beamspace directions, $\vbf_\ell$
and $\ubf_\ell$.  The coefficient $h_{k\ell d}$ is a scalar small-scale fading
coefficient that may vary with the subsignal $d$, transmission $\ell$ and scan cycle $k$.
Of course, real channels are composed of multiple path clusters with beamspread,
none of which will align exactly with one of the directions.
However, we will use this
idealized beamspace model only to simplify the detector design and analysis.
Our simulations, on the other hand, will be performed using a measurement-based realistic channel model.

Now, if we assume that the beamforming directions are orthonormal, i.e., $
    \ubf_{\ell}^*\ubf_{\ell_0}\vbf_{\ell}^*\vbf_{\ell_0} = \delta_{\ell,\ell_0},$
we can rewrite \eqref{eq:ruv} as
\beq \label{eq:ruvalpha}
    \rbf_{k\ell d} = \psi_{kd} \delta_{\ell,\ell_0}\sbf_{\ell d} + \wbf_{k\ell d},
\eeq
where $\psi_{kd} = h_{k\ell_0 d}$ and $\ell_0$ is the true direction.
The variable $\psi_{kd}$ is the complex gain
on the $d$-th subsignal in the $k$-th scan cycle at the time when the beam directions
are properly aligned.

The detection problem can then be posed as a hypothesis testing problem.
Specifically, the presence or absence of the signal corresponds to two hypotheses:
\beq \label{eq:hyp}
    \begin{array}{ll}
    H_0: ~\psi_{kd} = 0& \mbox{(signal absent),} \\
    H_1: ~\psi_{kd} \neq 0 & \mbox{(signal present).}
    \end{array}
\eeq
Now, the model in \eqref{eq:ruvalpha}
specifies the probability distribution of the observed data,
which we denote as
\beq
    p(\rbf|\taubf,\psibf,\ell_0),
\eeq
where $\rbf = \{\rbf_{k\ell d}\}$ is the set of all measurements,
$\taubf=\{\tau_{k\ell d}\}$ is
the set of noise levels, $\psibf=\{\psi_{kd}\}$ is the set of signal levels.
Since the model has unknown parameters, we follow
the procedure in \cite{BarHosCellSearch:TWC15} and use a standard Generalized Likelihood
Ratio Test (GLRT)  \cite{VanTrees:01a} to decide between the two hypothesis.
Specifically, we first compute the minimum negative log
likelihood (ML) of each hypothesis,
\begin{subequations} \label{eq:MLhyp}
\beqa
    && \Lambda_0 := \min_{\taubf,\ell_0} -\ln p(\rbf|\taubf,\psibf = 0,\ell_0) \\
    && \Lambda_1 := \min_{\taubf,\psibf,\ell_0} -
    \ln p(\rbf|\taubf,\psibf,\ell_0),
\eeqa
\end{subequations}
and then use the test
\beq \label{eq:hypTest}
    \Hhat = \begin{cases}
        1 & \Lambda_0 - \Lambda_1 \geq t \\
        0 & \Lambda_0 - \Lambda_1 < t,
        \end{cases}
\eeq
where $t$ is a threshold.  It is shown in Appendix~\ref{sec:GLRTderiv} that the
test can be computed as follows. 
For each transmission $\ell$ in scan cycle $k$,
we compute the correlation
\beq\label{eq:rhokl}
    \rho_{k\ell d} := \frac{|\sbf_{\ell d}^*\rbf_{k\ell d}|^2}{
        \|\sbf_{\ell d}\|^2\|\rbf_{k\ell d}\|^2}.
\eeq
This is simply a matched filter.  We next compute the optimal beamspace direction
by minimizing
\beq \label{eq:ellhat}
    \ellhat_0 = \argmin_{\ell=1,\ldots,L}  \sum_{k=1}^K \sum_{d=1}^{N_{\rm div}} \ln\left(1-\rho_{k\ell d}\right).
\eeq
Then, it is shown in Appendix~\ref{sec:GLRTderiv} that the log likelihood difference is given by
\beq \label{eq:Trho}
    \Lambda_0-\Lambda_1 = M\sum_{k=1}^K \sum_{d=1}^{N_{\rm div}}
    \ln\left(1-\rho_{k\ellhat_0 d}\right),
\eeq
which can be applied into the hypothesis test \eqref{eq:hypTest}.

\subsection{Hybrid and Digital Beamforming}\label{sec:hybDig}

The above detector was derived for the case of analog BF.
However, the same detector can be used for hybrid and digital BF.
In hybrid BF at the RX, the RX has $S$ complete RF chains 
and thus has the ability to operate as
$S$ analog BF systems in parallel.
Thus, an identical detector can be used. The only difference is that one
can obtain $S$ power measurements in each time instant -- hence speeding up the
detection by a factor of $S$.

For digital BF, one can simply test all
$N_{\rm rx}$ directions in each time step,
which is again equivalent to analog BF but with an even faster acceleration.
Note that in digital BF, one can also test angles of arrival that are not
exactly along one of the $N_{\rm rx}$ beamspace directions. 
However, to keep the analysis of various design options simple we will not 
consider this detector but assume that in digital BF
the arrival angles are aligned with the beamspace directions.

Of course, as discussed above, digital BF may require much higher power
consumption to support separate ADCs for each antenna element.
In the analysis below, we will thus consider a
low-resolution digital architecture.

\section{Delay Analysis} \label{sec:delAn}

Given the search algorithm, we can now evaluate each of the
design options described in Table~\ref{tab:RachOpCombo}.  Our goal is to
determine two key delay
parameters: (1) The \emph{synchronization delay}, i.e., the
time it takes the UE to detect the presence of the synchronization
signal in Step 0 of Figure~\ref{fig:initAccess}, and (2)
the \emph{RA preamble detection delay}, i.e., the time the BS
needs to reliably detect the random access preamble in Step 1.
These first two steps will be the dominant components of the
overall delay for initial access.
We call the sum of the synchronization and RA delay the
\emph{access delay}.
Note that this delay is the baseline for the total, control plus data plane, latency.
Due to beam ``fine-tuning'' and channel estimation delay before each transmission in data plane,
a subject discussed in \cite{shokri2015beamsearch} and \cite{ bogale2015hybridADestimation},
it is crucial to push the access delay as low as possible in order to achieve the
very high throughput and low latency targets set for 5G \cite{andrews2014will}. 


Before undertaking a detailed numerical simulation, 
it is useful to perform a simple theoretical analysis that illustrates 
the fundamental relations between beamforming gains 
and delay for the different design options.
We focus on the synchronization phase which, since both the UE and the BS don't know the correct directions, is more challenging than RA.
To this end, suppose the initial access system targets 
all UEs whose SNR is above some minimum target value, $\gamma_{\rm tgt}$.
This target SNR may be defined for a target rate the UE is assumed to require
when in connected mode, and should be measured relative
to the SNR that a UE could get with beamforming gains, i.e.,
\beq \label{eq:gamtgt}
	\gamma_{\rm tgt} = \frac{PG_{\rm rx}G_{\rm tx}}{N_0 W_{\rm tot}},
\eeq
where $P$ is the omni-directional received power (affected by the pathloss), $W_{\rm tot}$ 
the total available bandwidth in connected mode, and $G_{\rm rx}$ and $G_{\rm tx}$ are the maximum RX and TX antenna gains,
i.e., the gains that the link could enjoy if the directions were perfectly aligned.

Now, let $\gamma_{\rm sig}$ be the minimum SNR accumulated on the synchronization signal required for reliable detection (the precise value will depend on the false alarm and misdetection probabilities). Each synchronization transmission will have an omni-directional 
received energy of $PT_{\rm sig}$.   
In all design options in the synchronization phase, the UE scans the space directionally, 
obtaining a maximum gain of $G_{\rm rx}$ when the beams are aligned.
We let $G_{\rm tx}^{\rm sync}$ be the maximum TX beamforming gain available in the 
synchronization phase:  
$G_{\rm tx}^{\rm sync}=1$ for the omni-directional transmission 
and $G_{\rm tx}^{\rm sync}=G_{\rm tx}$ for the case of directional transmissions.
Thus, when the beams are correctly aligned, the synchronization signal will 
have a received energy of $ G_{\rm rx}G_{\rm tx}^{\rm sync}PT_{\rm sig}.$
The beams will be correctly aligned exactly once every scan cycle, and hence, to obtain reliable detection after $K$ scan cycles we need
\beq \label{eq:gamsigbnd}
	\frac{KPT_{\rm sig,sync}G_{\rm rx}G_{\rm tx}^{\rm sync}}{N_0} \geq \gamma_{\rm sig}.
\eeq  
Following Fig.~\ref{fig:acsSig}, the synchronization signal occupies $T_{\rm sig,sync}$ seconds
every $T_{\rm per,sync}$ seconds and hence the overhead is
\beq \label{eq:ovSync}
    \phi_{\rm ov,sync} = T_{\rm sig,sync}/T_{\rm per,sync},
\eeq 
and the time to perform $K$ scan cycles in
\beq \label{eq:dklt}
	D_{\rm sync} = KLT_{\rm per,sync}.
\eeq
Combining \eqref{eq:gamtgt}, \eqref{eq:gamsigbnd} and \eqref{eq:ovSync},
the synchronization delay $D_{\rm sync}$ can be bounded as
\begin{align}
	D_{\rm sync} &= KLT_{\rm per,sync} \geq \frac{\gamma_{\rm sig}LG_{\rm tx}}
    	{G_{\rm tx}^{\rm sync}\gamma_{\rm tgt}W_{\rm tot}\phi_{\rm ov,sync}}.
\end{align}
Now, the antenna gain is proportional to the number of antennas, and we will thus 
assume that the maximum gain is given by
$
	G_{\rm rx} = N_{\rm rx}, \quad 	G_{\rm tx} = N_{\rm tx}.
$
Also, for the analog BF options, ODx and DDx, it can be verified that
$L=G_{\rm rx}G_{\rm tx}^{\rm sync}$, so
\beq
	D_{\rm sync}  \geq \frac{\gamma_{\rm sig}G_{\rm rx}G_{\rm tx}}
    	{\gamma_{\rm tgt}W_{\rm tot}\phi_{\rm ov,sync}}.  \label{eq:Dsbnd1}
\eeq

This bound may not be achievable, since it may require that
the synchronization signal duration $T_{\rm sig,sync}$ is very small.
Very short synchronization signals may not be practical since the signal
needs sufficient degrees of freedom to accurately estimate the noise in addition 
to the signal power -- we will see this effect in the numerical simulations below.
Thus, suppose there is some minimum practical synchronization signal time, 
$T_{\rm sig,sync}^{\rm min}$.  Since $K \geq 1$,
\begin{align}
	D_{\rm sync} &= KLT_{\rm per} \geq \frac{L T_{\rm sig,sync}}{\phi_{\rm ov,sync}} \nonumber \\
    	&\geq 
    	\frac{L T_{\rm sig,sync}^{\rm min}}{\phi_{\rm ov,sync}} =
        \frac{G_{\rm rx}G_{\rm tx}^{\rm sync}T_{\rm sig,sync}^{\rm min}}{\phi_{\rm ov,sync}}, \label{eq:Dsbnd2}
\end{align}
Combining \eqref{eq:Dsbnd1} and \eqref{eq:Dsbnd2},
\begin{align}
	D_{\rm sync} \geq \frac{G_{\rm rx}}{\phi_{\rm ov,sync}}\max\left\{
    	\frac{\gamma_{\rm sig}G_{\rm tx}}
    	{\gamma_{\rm tgt}W_{\rm tot}}, 
        G_{\rm tx}^{\rm sync}T_{\rm sig,sync}^{\rm min}
        \right\}. \label{eq:Dsbndana}
\end{align}
The digital case is similar except that $L=G_{\rm tx}^{\rm sync}$ since only the transmitter needs to sweep the different directions.
Hence, the synchronization delay is reduced by a factor of $G_{\rm rx}$:
\begin{align}
	D_{\rm sync} \geq \frac{1}{\phi_{ov,sync}}\max\left\{
    	\frac{\gamma_{\rm sig}G_{\rm tx}}
    	{\gamma_{\rm tgt}W_{\rm tot}}, 
        G_{\rm tx}^{\rm sync}T_{\rm sig,sync}^{\rm min}
        \right\}. \label{eq:Dsbnddig}
\end{align}

This simple analysis reveals several important features.
First consider the delay under analog BF in \eqref{eq:Dsbndana}.
Without a bound on $T_{\rm sig,sync}$, both analog BF options ODx and DDx will have the same delay, and the delay grows linearly with the combined gain $G_{\rm rx}G_{\rm tx}$.
Hence, there is a direct relation between 
delay and the amount of BF gain used in the system.  
However, when using a signal duration equal to the bound $T_{\rm sig, sync}^{\rm min}$, the second term in the maximum in \eqref{eq:Dsbndana} will become dominant. 
In this case, an omni directional transmission will have a lower bound delay of only $T_{\rm sig, sync }^{\rm min}$ (since $G^{\rm sync}_{\rm tx} = 1$), while a directional one will be lower-bounded by $G^{\rm sync}_{\rm tx} T_{\rm sig, sync }^{\rm min} > T_{\rm sig, sync }^{\rm min}$.

Finally, digital BF at the UE (Option DDigX) offers a significant 
improvement over both analog options by a factor of $G_{\rm rx}$.
Hence, in circumstances where the UE has a high directional gain antenna,
the benefits can be significant.

\section{Numerical Evaluation} \label{sec:evalan}

\subsection{System Parameters}
The above analysis is highly simplified and, while useful to gain some intuition
and basic understanding, cannot be used to make
detailed system design choices.   
For a more accurate analysis of the delay,
we assessed the delay under realistic system parameters and channel models.
Appendix~\ref{sec:simParam} provides a detailed discussion on the reasoning behind the parameter selection.
In brief, the signal parameters $T_{\rm sig}$ and $W_{\rm sig}$
were selected to ensure that the channel is roughly flat within each
$T_{\rm sig} \x W_{\rm sig}$ subsignal
time-frequency region based on the typical values of
the coherence time and the coherence bandwidth observed in
\cite{Rappaport:12-28G,McCartRapICC15}.
The parameter $T_{\rm sig}$ was also selected sufficiently short so that
the Doppler shift across the subsignal would not be significant
with moderate UE velocities (30~km/h).
Depending on the overhead that we can tolerate for synchronization and random access signals, different values for $T_{\rm per}$ are selected.
The use of $N_{\rm div}=4$ subsignals was found experimentally to give the
best performance in terms of frequency diversity versus energy loss from
non-coherent combining. For the antenna arrays, we followed the capacity
analysis in \cite{AkdenizCapacity:14} and
considered 2D uniform planar arrays with $4 \times 4$
elements at the UE, and $8 \times 8$ at the BS.

To compute the threshold level $t$ in \eqref{eq:Trho},
we first computed a false alarm probability target with the formula
\[
    P_{\textsc FA} = \frac{R_{\textsc FA}}{N_{\rm sig}N_{\rm dly}N_{\textsc FO}},
\]
where $R_{\textsc FA}$ is the maximum false alarm rate per search period over
all signal, delay and frequency offset hypotheses and $N_{\rm sig}$, $N_{\rm dly}$
and $N_{\textsc FO}$ are, respectively, the number of signal,
delay and frequency offset hypotheses. 
Again, the details of the selection are given
in Appendix~\ref{sec:simParam}. $N_{\rm dly}$ represents the number of delay hypotheses in each transmission period in either uplink or downlink.
The detector has to decide at which delay $\tau$ the signal was received.
The granularity of searching for the correct $\tau$ is at the level of one out of all the samples between two transmission periods $T_{\rm per}$.
Assuming sampling at twice the bandwidth, in the downlink direction,
this is $\frac{1}{2W_{\rm sig,sync} \times T_{\rm per,sync}}$.
Hence, the number of delay hypotheses, $N_{\rm dly}$, is equal to all
these possible $\tau$s. 
In other words, $N_{\rm dly} = 2W_{\rm sig,sync}T_{\rm per,sync}$.
Note that, since the signal is transmitted periodically, 
$N_{\rm dly}$ stays constant and does not increase as time passes.
In the uplink direction we have
$N_{\rm dly} = 2W_{\rm sig,RA}T_{\rm offset}$,
where $T_{\rm offset}$ is the round trip time of the
signal propagation between UE and BS (See Appendix B for why this number is different for RA and Sync.)
The number of frequency offset hypotheses $N_{\textsc FO}$ was computed
so that the frequency search could cover both an
initial local oscillator (LO) error of $\pm 1$~part per million (ppm) as well as a Doppler shift
up to 30~km/h at 28~GHz.
We selected
$N_{\rm sig}=3$ for the number of signal hypotheses in the synchronization step and $N_{\rm sig}=64$ for the random access step as used in current 3GPP LTE.
We then used a large number of Monte Carlo trials to find the
threshold to meet the false alarm rate.
Having set this target $P_{FA}$, we essentially wish to find the correlation $\rho_{k \ell d}$ at point $t$ which corresponds to this probability.
This point will be the detection threshold \cite{VanTrees:01a}.
As long as $\rho_{k \ell d}$ is below this threshold the detector assumes that just noise is being received, while if $\rho_{k \ell d}$ goes above it, the detector will assume $\rho_{k \ell d} = \rho_{k \hat{\ell}_0 d}$ and hence an estimate of the correct angle $\ell_0$ is found.
As shown in Appendix~\ref{sec:simParam} for our selected parameters, $P_{\textsc FA}$ becomes very small.
Therefore, we extrapolated the tail distribution of the statistic
$T = \rho_{k \ell d}$ of the test defined by \eqref{eq:hypTest} and \eqref{eq:Trho}, to estimate the threshold analytically.
To this end we used a second degree polynomial fit for $\log \Pr(T > t)$ \cite{BarHosCellSearch:TWC15}.

\ifonecol

\begin{table}
\centering
\setlength\tabcolsep{4pt}
\begin{minipage}[t]{0.49\textwidth}
\centering
\begin{tabular}{|p{2.7cm}|p{4.3cm}|}
  \hline
  {\bf Parameter} & {\bf Value}
  \tabularnewline \hline

Cell radius, $r$ & $100$~m
\tabularnewline \hline

Downlink TX power & $30$~dBm
\tabularnewline \hline

Uplink TX power & $20$~dBm
\tabularnewline \hline

BS noise figure, $NF_{BS}$ & $4$~dB
\tabularnewline \hline

UE noise figure, $NF_{UE}$  & $7$~dB
\tabularnewline \hline

Thermal noise power density, $kT$  & $-174$~dBm/Hz
\tabularnewline \hline

Carrier Frequency & 28 GHz
\tabularnewline \hline

Total system bandwidth, $W_{\rm tot}$ & 1 GHz
\tabularnewline \hline

Probability of LOS vs.\ NLOS & $P_{LOS}(d)=\exp (-a_{los} d)$, $ a_{los}= 67.1$~m
\tabularnewline \hline

Path loss model dB & $PL(d) =  \alpha + 10 \beta \log10(d) +\xi$  $\xi \sim \mathcal{N} (0, \sigma^2)$,
$d$ is in meters
\tabularnewline \hline

NLOS parameters & $\alpha=72.0$, $\beta=2.92$, $\sigma=8.7$ dB \tabularnewline \hline
LOS parameters & $\alpha=61.4$, $\beta=2.0$, $\sigma=5.8$ dB \tabularnewline \hline
\end{tabular}
\end{minipage}%
\hfill
\setlength\tabcolsep{4pt}
\begin{minipage}[t]{0.49\textwidth}
\centering

\begin{tabular}{|p{3.5cm}|p{4.1cm}|}
  \hline
  {\bf Parameter} & {\bf Value}
  \tabularnewline \hline


Number of subsignals per time slot, $N_{\rm div}$    & $4$
\tabularnewline \hline

Subsignal Duration, $T_{\rm sig}$ & varied  ($10$ $\mu$s, $50$ $\mu$s, $100$ $\mu$s)
\tabularnewline \hline

Subsignal Bandwidth, $W_{\rm sig}$ & 1 MHz
\tabularnewline \hline

Period between transmissions, $T_{\rm per}$  & Varied to meet overhead requirements
\tabularnewline \hline


Total false alarm rate per scan cycle,
$R_{\textsc FA}$ & $ 0.01$
\tabularnewline \hline

Number of sync signal waveform hypotheses  & $3$
\tabularnewline \hline

Number of RA preambles & $64$ \tabularnewline \hline

Number of frequency offset hypotheses, $N_{\textsc FO}$ & $ 23 $ \tabularnewline \hline

BS antenna & $8\times8$ uniform planar array   \tabularnewline \hline
UE antenna & $4\times4$ uniform planar array   \tabularnewline \hline



Number of bits in the low resolution digital architecture & 3 per I/Q \tabularnewline \hline

\end{tabular}
\end{minipage}\\[7pt]
\begin{minipage}[t]{.49\linewidth}
        \caption{SNR simulation parameters.}
        \label{tab:simSNRpar}
    \end{minipage}%
    \hfill%
    \begin{minipage}[t]{.49\linewidth}
        \caption{Default simulation parameters unless otherwise stated. Carrier frequency, cell radius and total system bandwidth are the same as in Table~\ref{tab:simSNRpar}.}
     \label{tab:simPar}
    \end{minipage}%

\end{table}

\else

\begin{table}
\begin{center}
\ifonecol
\begin{tabular}{|>{\raggedright}p{6 cm}|>{\raggedright}p{5 cm}|}
\else
\begin{tabular}{|>{\raggedright}p{3.25 cm}|>{\raggedright}p{3 cm}|}
\fi
  \hline
  {\bf Parameter} & {\bf Value}
  \tabularnewline \hline

Total system bandwidth, $W_{\rm tot}$ & 1 GHz
\tabularnewline \hline

Number of subsignals per time slot, $N_{\rm div}$    & $4$
\tabularnewline \hline

Subsignal Duration, $T_{\rm sig}$ & varied  ($10$ $\mu$s, $50$ $\mu$s, $100$ $\mu$s)
\tabularnewline \hline

Subsignal Bandwidth, $W_{\rm sig}$ & 1 MHz
\tabularnewline \hline

Period between transmissions, $T_{\rm per}$  & Varied to meet overhead requirements
\tabularnewline \hline


Total false alarm rate per scan cycle,
$R_{\textsc FA}$ & $ 0.01$
\tabularnewline \hline

Number of sync signal waveform hypotheses  & $3$
\tabularnewline \hline

Number of RA preambles & $64$ \tabularnewline \hline

Number of frequency offset hypotheses, $N_{\textsc FO}$ & $ 23 $ \tabularnewline \hline

BS antenna & $8\times8$ uniform planar array   \tabularnewline \hline
UE antenna & $4\times4$ uniform planar array   \tabularnewline \hline

Carrier Frequency & 28 GHz \tabularnewline \hline

Cell radius & 100~m \tabularnewline \hline

Number of bits in the low resolution digital architecture & 3 per I/Q \tabularnewline \hline

\end{tabular}
\caption{Default simulation parameters unless otherwise stated.}
\label{tab:simPar}
\end{center}
\end{table}

\begin{table}
\begin{center}
\ifonecol
\begin{tabular}{|>{\raggedright}p{6 cm}|>{\raggedright}p{5 cm}|}
\else
\begin{tabular}{|>{\raggedright}p{3 cm}|>{\raggedright}p{4 cm}|}
\fi
  \hline
  {\bf Parameter} & {\bf Value}
  \tabularnewline \hline

Cell radius, $r$ & $100$~m
\tabularnewline \hline

Downlink TX power & $30$~dBm
\tabularnewline \hline

Uplink TX power & $20$~dBm
\tabularnewline \hline

BS noise figure, $NF_{BS}$ & $4$~dB
\tabularnewline \hline

UE noise figure, $NF_{UE}$  & $7$~dB
\tabularnewline \hline

Thermal noise power density, $kT$  & $-174$~dBm/Hz
\tabularnewline \hline

Carrier Frequency & 28 GHz
\tabularnewline \hline

Total system bandwidth, $W_{\rm tot}$ & 1 GHz
\tabularnewline \hline

Probability of LOS vs.\ NLOS & $P_{LOS}(d)=\exp (-a_{los} d)$, $ a_{los}= 67.1$~m
\tabularnewline \hline

Path loss model dB & $PL(d) =  \alpha + 10 \beta \log10(d) +\xi$  $\xi \sim \mathcal{N} (0, \sigma^2)$,
$d$ is in meters
\tabularnewline \hline

NLOS parameters & $\alpha=72.0$, $\beta=2.92$, $\sigma=8.7$ dB \tabularnewline \hline
LOS parameters & $\alpha=61.4$, $\beta=2.0$, $\sigma=5.8$ dB \tabularnewline \hline
\end{tabular}
\caption{SNR simulation parameters.}
\label{tab:simSNRpar}
\end{center}
\end{table}

\fi

\subsection{SNR} \label{sec:simSNR}

We used the following model for the SNR distribution.
We envision a cell area of radius $r = 100$~m with the mmWave BS at its center, and  UEs randomly ``dropped'' within this area.
We then compute a random path loss between the BS and the UEs based on the model
in \cite{AkdenizCapacity:14}.
This model was obtained from data gathered in multiple measurement campaigns in downtown
New York City \cite{Rappaport:12-28G,Rappaport:28NYCPenetrationLoss,Samimi:AoAD,rappaportmillimeter,McCartRapICC15}.
The path loss model parameters, along with the rest of the simulation parameters,
are given in Table \ref{tab:simSNRpar}.

In the model in \cite{AkdenizCapacity:14}, the channel/link between a UE and a BS is randomly selected, according to some distance dependent probabilities, to be in one of three states, namely, LOS, NLOS or outage. Here, we ignore the outage state, since UEs in outage cannot establish a link, and instead focus on the other two states, in which initial access can be performed.
A UE may be in either state with probabilities $P_{\textsc LOS}(d)$ or $P_{\textsc NLOS} = 1-P_{\textsc LOS}(d)$.
The omni-directional path loss is then computed from
\[
    PL =  \alpha + 10 \beta \log10(d) +\xi \;[dB] , \; \xi \sim \mathcal{N} (0, \sigma^2),
\]
where the parameters $\alpha$, $\beta$ and $\sigma$ depend on the LOS or NLOS condition
and $d$ is the distance between the transmitter and the receiver.

We do not consider interference from other cells.
In the downlink, since the bandwidth in mmWave cells is large,
we assume that neighboring cells can be time
synchronized and place their synchronization signals on non-overlapping frequencies.
Since we assume that there are no other signals transmitted during the synchronization signal slot
(see Fig.~\ref{fig:acsSig}), using partial frequency reuse does not increase the overhead.
In the uplink, we ignore collisions from other random access transmissions either in-cell
or out-of-cell, so again we can ignore the interference.

\begin{figure}
\centering {\includegraphics[width=0.9\linewidth]{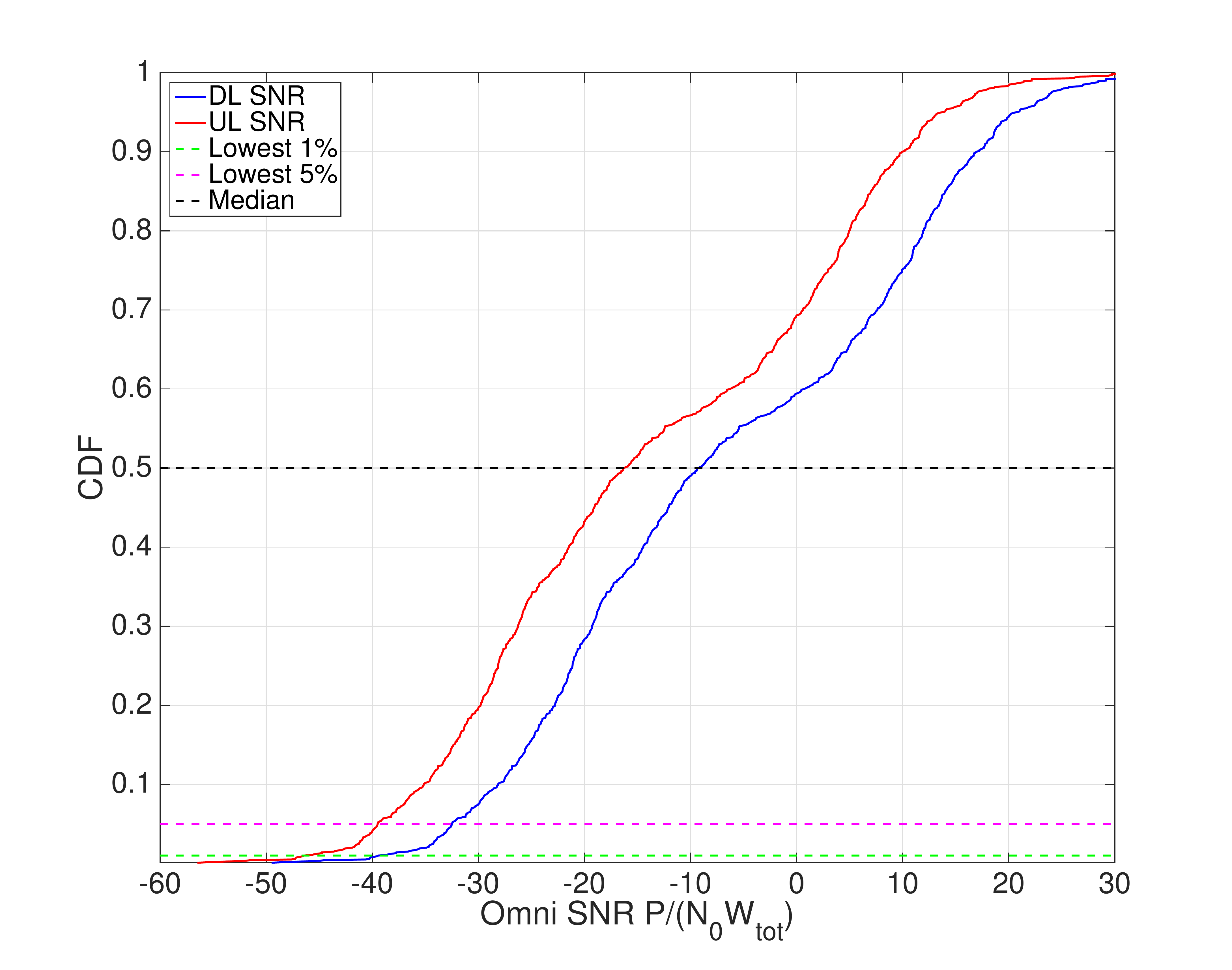}}
\caption{SNR distribution in downlink and uplink for a cell of radius 100 m.}
\label{fig:snrdistr}
\end{figure}

Fig.~\ref{fig:snrdistr} shows the DL and UL SNR distributions under these
assumptions.  The SNR plotted is measured on a total bandwidth of $W_{\rm tot}=1$ GHz,
\beq \label{eq:SNRtot}
    \SNR_{\rm data,omni} = \frac{P}{N_0W_{\rm tot}},
\eeq
where $P$ is the received power and $N_0$ is the noise power spectral density (including the noise figure).
Note that this is the \emph{omni-directional SNR} and hence does not include the beamforming gain.

The synchronization and RA signals are transmitted
in much narrower bands and therefore will have higher
SNRs in the signal bandwidth.
The reason we plot the SNR with respect to the total bandwidth
is that we believe it makes it easier to interpret
the results of Fig.~\ref{fig:snrdistr}.
Plotting the narrowband RA/Sync SNRs would be somehow irrelevant since they are meaningless beyond the context of these very signals. 
It is the SNR over the whole available bandwidth $W_{\rm tot}$ that defines the SNR
regime where a UE operates. It is this SNR that determines the data rates in both DL and UL -- See Appendix~\ref{sec:simParam} for the connection between $\SNR_{\rm data,omni}$ and the signals' SNRs.
Also plotted are the $1\%$ and $5\%$ percentiles and the median lines for the UL and DL.

\subsection{Fully Digital BF with Low Bit Resolution}\label{sec:dig_lowbit}
Two of the options in Table~\ref{tab:RachOpCombo} require fully digital receivers:
ODDig requires a fully digital front-end at the BS and ODigDig requires one at both the UE and the BS.
Deploying such a fully digital front end comes at the potential 
cost of high power consumption due to the need for one analog to digital converter
(ADC) per antenna element.
This problem is especially of concern on the UE side where the power consumption of the device 
should remain as low as possible.

\begin{figure}
\centering {\includegraphics[width=0.95\linewidth]{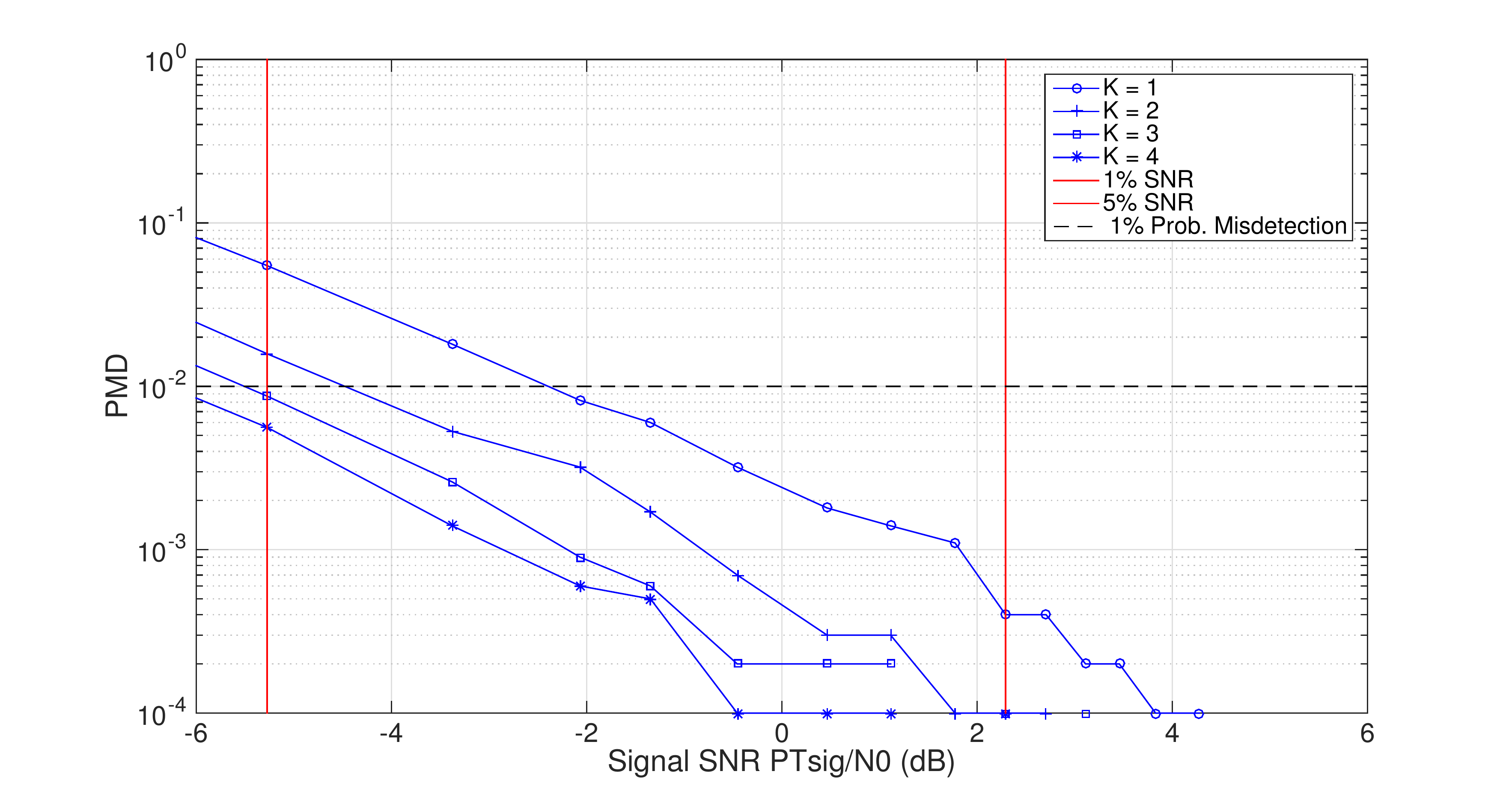}}
\caption{	DDD/DDO, probability of misdetection (PMD) versus synchronization signal SNR for $K = 1, 2, 3, 4$.
Each additional cycle decreases the misdetection probability for a given SNR regime. }
\label{fig:sync_pmd}
\end{figure}

\ifthreebitq
To properly compare fully digital options with the analog BF cases, we constrain the digital 
architecture to have the \emph{power consumption in the same order as the analog BF} case by using
a very small number of bits per I/Q dimension.
Such low bit resolution mmWave front-ends have been studied in 
\cite{Madhow:ADC,Madhow:largeArray,mo2014high,mo2014channel,OrhanITA15}.
In \cite{OrhanITA15} specifically, the authors discuss among other things the impact of ADC 
resolution on the communication rate after the link has been established. 
The concept is that the
power consumption of an ADC generally scales as $2^b$, 
where $b$ is the number of bits ~\cite{razavi2001design}. 
In the analysis below, we will constrain the fully digital architecture 
to have $b=3$ bits per I/Q dimension at most. 
Since conventional ADCs typically have near 10 bits, 
using 3 bits should reduce the power consumption by a factor of 128 which more 
than compensates for adding an ADC on all 16 or 64 elements of the antenna array.

\begin{table}
\centering
\begin{tabular}{|p{2cm}|p{5cm}|}
  \hline
  {\bf Number bits} & {\bf Quantization loss, $-10 \log_{10} \frac{\gamma_{\rm hq}}{\gamma_{\rm lq}}$ (dB)}
  \tabularnewline \hline
  1 &   1.96 \tabularnewline \hline
  2 &   0.54 \tabularnewline \hline
  3 &   0.15 \tabularnewline \hline
\end{tabular}
\caption{Quantization loss as a function of the number of bits for
a scalar uniform quantizer with Gaussian noise.
\label{tbl:quant} }
\end{table}

\fi

To analyze the effect of the low resolution, we use a standard additive noise quantization model
\cite{GershoG:92,FletcherRGR:07} as used in the analysis of fully digital receivers for
cell search in \cite{BarHosCellSearch:TWC15}.
There it is shown that the effective SNR due to the 
quantization noise is given by
\beq \label{eq:gammalq}
    \gamma_{\rm lq} = \frac{(1-\sigma)\gamma_{\rm hq}}{1+\sigma\gamma_{\rm hq}},
\eeq
where $\gamma_{\rm lq}$ and $\gamma_{\rm hq}$ are the effective SNR of the low resolution ADC,
and the initial SNR of the high resolution ADC, respectively. The factor $\sigma$ is 
the average relative error of the quantizer and depends on the quantizer design,
input distribution and number of bits.  Equation~\eqref{eq:gammalq} 
shows that at very low SNRs,
the quantization results in a loss of $1-\sigma$.  
\ifthreebitq
Table~\ref{tbl:quant} shows this loss for $b=1$ to 3 bits using the numbers from
\cite{BarHosCellSearch:TWC15}.  
We see that at $b=3$, the quantization loss is only 0.15~dB.  
In all the subsequent analysis below, we will restrict the digital solutions to have only $b=3$ bits
and use \eqref{eq:gammalq} to transform the SNRs to account for the increased quantization noise.
\fi


\ifthreebitq

\ifonecol
\begin{table}
\begin{center}
\begin{tabular}{ |c|c|c|c|c|c|c|c|c|c|c|c|c|c|c|c|}
 \hline
 \multirow{3}{4em}{Option}&\multirow{3}{4em}{$T_{\rm sig}$} & \multicolumn{2}{|c|}{$L$} & \multicolumn{4}{|c|}{1\% SNR} & \multicolumn{4}{|c|}{5\% SNR} & \multicolumn{4}{|c|}{High SNR} \\
&& Sync & RA & \multicolumn{2}{|c|}{Sync} & \multicolumn{2}{|c|}{RA} & \multicolumn{2}{|c|}{Sync} & \multicolumn{2}{|c|}{RA} & \multicolumn{2}{|c|}{Sync} & \multicolumn{2}{|c|}{RA} \\
&&&&$K^*$&delay&$K^*$&delay&$K^*$&delay&$K^*$&delay&$K^*$&delay&$K^*$&delay\\
\hline
 \multirow{3}{4em}{DDO} & $10~\mu$s & & &3&614.4&2393&478.6 &1&204.8&65&13&1&204.8&1&0.2 \\
 &$50~\mu$s &1024 &1 &1&1024 &60 & 60 &1&1024& 3&3 &1&1024&1& 1 \\
&$100~\mu$s & & &1&2048&17& 34 &1&2048& 2&4 &1&2048&1& 2\\
\hline
\multirow{3}{4em}{DDD} & $10~\mu$s & & &3&614.4& 24& 307.2 &1&204.8&2&25.6&1&204.8&1&12.8\\
 &$50~\mu$s &1024 & 64 &1&1024 &2& 128 &1&1024&1&64&1&1024&1&64\\
&$100~\mu$s & & &1&2048&1&128&1&2048&1&128&1&2048&1&128\\
\hline
 \multirow{3}{4em}{ODD} & $10~\mu$s & & & 73&233.6 &24& 307.2 &4&12.8&2&25.6 &1&3.2&1&12.8 \\
 &$50~\mu$s &16 & 64 & 3&48&2&128&1&16&1&64&1&16&1&64\\
&$100~\mu$s & & &1&32&1&128&1&32&1&128&1&32&1&128\\
\hline
 \multirow{3}{4em}{ODDig} & $10~\mu$s & & &  73&233.6  &25& 5 &4&12.8&2&0.4 &1&3.2&1&0.2\\
 &$50~\mu$s &16 &1 & 3&48&2&2&1&16&1&1&1&16&1&1\\
&$100~\mu$s & & &1&32&1&2&1&32&1&2&1&32&1&2 \\
 \hline
 \multirow{3}{4em}{ODigDig} &$10~\mu$s & & &  79&15.8  &25& 5  &4&0.8  &2&0.4 &1&0.2 &1&0.2 \\
 &$50~\mu$s &1 &1 &4& 4&2&2 &1& 1&1&1 &1&1 &1&1 \\
&$100~\mu$s & & & 2&4&1&2&1& 2&1& 2&1& 2&1&2 \\
 \hline
\end{tabular}
\caption{ Detection delays in milliseconds for a fixed overhead of 5$\%$.
The third and fourth column show the size of the angular space $L$ that has to be scanned in Sync and RA for each option. Note that they are the same as the fifth and sixth columns of Table~\ref{tab:simPar} except that we have assigned numbers to $N_{\rm tx}$ and $N_{\rm rx}$. $K^*$ is the number of full cycles required for detection in each scenario. Values for digital BF are derived given an ADC with 3 bit resolution.}
\label{tab:delays}
\end{center}
\end{table}

\else

\begin{table*}
\begin{center}
\begin{tabular}{ |c|c|c|c|c|c|c|c|c|c|c|c|c|c|c|c|}
 \hline
 \multirow{3}{4em}{Option}&\multirow{3}{4em}{$T_{\rm sig}$} & \multicolumn{2}{|c|}{$L$} & \multicolumn{4}{|c|}{1\% SNR} & \multicolumn{4}{|c|}{5\% SNR} & \multicolumn{4}{|c|}{High SNR} \\
&& Sync & RA & \multicolumn{2}{|c|}{Sync} & \multicolumn{2}{|c|}{RA} & \multicolumn{2}{|c|}{Sync} & \multicolumn{2}{|c|}{RA} & \multicolumn{2}{|c|}{Sync} & \multicolumn{2}{|c|}{RA} \\
&&&&$K^*$&delay&$K^*$&delay&$K^*$&delay&$K^*$&delay&$K^*$&delay&$K^*$&delay\\
\hline
 \multirow{3}{4em}{DDO} & $10~\mu$s & & &3&614.4&2393&478.6 &1&204.8&65&13&1&204.8&1&0.2 \\
 &$50~\mu$s &1024 &1 &1&1024 &60& 60 &1&1024& 3&3 &1&1024&1& 1 \\
&$100~\mu$s & & &1&2048&17& 34 &1&2048& 2&4 &1&2048&1& 2\\
\hline
\multirow{3}{4em}{DDD} & $10~\mu$s & & &3&614.4& 24& 307.2 &1&204.8&2&25.6&1&204.8&1&12.8\\
 &$50~\mu$s &1024 & 64 &1&1024 &2& 128 &1&1024&1&64&1&1024&1&64\\
&$100~\mu$s & & &1&2048&1&128&1&2048&1&128&1&2048&1&128\\
\hline
 \multirow{3}{4em}{ODD} & $10~\mu$s & & & 73&233.6 &24& 307.2 &4&12.8&2&25.6 &1&3.2&1&12.8 \\
 &$50~\mu$s &16 & 64 & 3&48&2&128&1&16&1&64&1&16&1&64\\
&$100~\mu$s & & &1&32&1&128&1&32&1&128&1&32&1&128\\
\hline
 \multirow{3}{4em}{ODDig} & $10~\mu$s & & &  73&233.6  &25& 5 &4&12.8&2&0.4 &1&3.2&1&0.2\\
 &$50~\mu$s &16 &1 & 3&48&2&2&1&16&1&1&1&16&1&1\\
&$100~\mu$s & & &1&32&1&2&1&32&1&2&1&32&1&2 \\
 \hline
 \multirow{3}{4em}{ODigDig} &$10~\mu$s & & &  79&15.8  &25& 5  &4&0.8  &2&0.4 &1&0.2 &1&0.2 \\
 &$50~\mu$s &1 &1 &4& 4&2&2 &1& 1&1&1 &1&1 &1&1 \\
&$100~\mu$s & & & 2&4&1&2&1& 2&1& 2&1& 2&1&2 \\
 \hline
\end{tabular}
\caption{ Detection delays in milliseconds for a fixed overhead of 5$\%$.
The third and fourth column show the size of the angular space $L$ that has to be scanned in Sync and RA for each option. Note that they are the same as the fifth and sixth columns of Table~\ref{tab:simPar} except that we have assigned numbers to $N_{\rm tx}$ and $N_{\rm rx}$. $K^*$ is the number of full cycles required for detection in each scenario. Values for digital BF are derived given an ADC with 3 bit resolution.}
\label{tab:delays}
\end{center}
\end{table*}
\fi
 
\fi

\subsection{Multipath NYC Channel Model} \label{sec:chanMod}
In deriving the GLRT detector in Section~\ref{sec:glrtDtct}
we assumed that the channel gain was
concentrated in a single path aligned exactly with one of the $L$
beamspace direction pairs.
This assumption may look naive at first, since real channel measurements 
have shown that the mmWave channel exhibits multipath characteristics \cite{Rappaport:12-28G,Rappaport:28NYCPenetrationLoss,Samimi:AoAD,rappaportmillimeter,McCartRapICC15}.
However, the fundamental difficulty in Sync/RA signal detection arises from the potentially large size of the angular space and the time needed to cover it.
Our detector sufficiently captures this difficulty for each and every one of the presented designs.
Moreover, in our previous work \cite{BarHosCellSearch:TWC15}, we showed that a multipath channel helps in detecting the desired signal as it provides more discovery opportunities through multiple macro-level scattering paths.
Nevertheless, while we will use this detector,
we simulate and evaluate our proposed designs against a realistic spatial 
non-line of sight (NLOS) channel model in 
\cite{AkdenizCapacity:14} derived from actual 28~GHz measurements 
in New York City~\cite{Rappaport:12-28G,Rappaport:28NYCPenetrationLoss,Samimi:AoAD,rappaportmillimeter,McCartRapICC15}. 
Due to limited space, we do not provide a detailed discussion on this channel model here, 
and just mention that it is described by multiple 
random clusters, with random azimuth and elevation angles and beamspread
-- see \cite{AkdenizCapacity:14} for details.
Finally, we note that the proposed analysis and methodology are general and, although used here in combination with the NYU measurement-based channel model to evaluate and compare the design options, can be applied to any other channel model or measurement data set.

\subsection{Synchronization Delay} \label{sec:syncDel}

We now proceed to evaluate the delays for each of the proposed
options in Table \ref{tab:RachOpCombo}.
We first estimate the number of scan cycles $K$ it takes
to reliably detect the synchronization signal from the BS.  This depends on 
the design option and the SNR.  
Figure \ref{fig:sync_pmd} shows the probability of misdetection ($P_{MD}$) as a function 
of the SNR and of the number of scan cycles $K$ for the DDD/DDO
design options, evaluated via on Monte Carlo simulations. 
Recall that the false alarm rate was set to $R_{\textsc FA} = 0.01$
per scan cycle. 
Also plotted are the vertical lines for the lowest $1\%$ and $5\%$ of the SNR DL distributions
described in Section~\ref{sec:simSNR}.
We set a target misdetection rate of $P_{MD}=0.01$ for a Sync signal of duration $T_{\rm sig,sync} = 10$ $\mu$s.
We observe in the same figure that the 1\% cell edge users, as defined by the SNR distribution obtained in Section~\ref{sec:simSNR}, will require $K=3$ scan cycles to meet the target while the 5\% and above users (according to the same distribution) are able to pass the detection target in just one scan cycle.  

The simulation is repeated for the other design options and different values of $T_{\rm sig,sync}$.
Based on the minimum number of required scan cycles, we can then compute the synchronization delay as a function of the overhead.
Following the analysis of Section~\ref{sec:delAn}, the relation between delay and overhead is
$
    D_{\rm sync} = KLT_{\rm per,sync} = \frac{KLT_{\rm sig,sync}}{\phi_{\rm ov,sync}}.
$
Using $K$ from the simulation, Table~\ref{tab:delays} shows the delays $D_{\rm sync}$ for a fixed overhead of $\phi_{\rm ov,sync}=$ 5\%.

The minimum synchronization signal time we consider is $T_{\rm sig,sync}^{min}=10~\mu$s.
Since each synchronization subsignal is transmitted over a narrow band $W_{\rm sig}$,
the degrees of freedom in each subsignal are $T_{\rm sig}W_{\rm sig}$.  
Since we have assumed $W_{\rm sig}=1$~MHz, using $T_{\rm sig,sync}=10~\mu$s would provide
10 degrees of freedom, and thus 9 degrees of freedom on which to estimate the noise variance.

\begin{figure}
\centering {\includegraphics[width=0.9\linewidth]{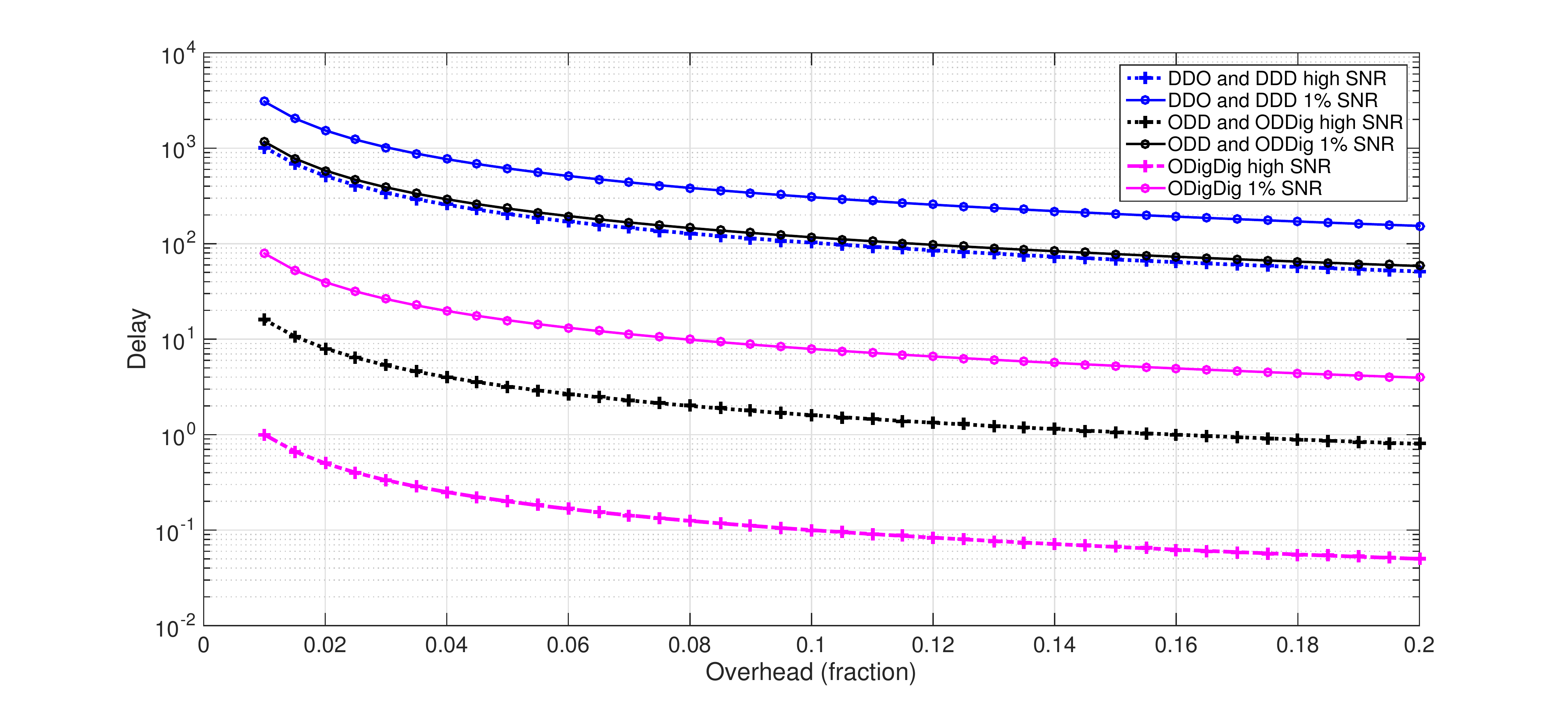}}
\caption{Synchronization delay in ms versus signal overhead for the minimum Sync signal duration $T_{\rm sig, sync}^{\rm min} = 10$~${\rm \mu}$s.}
\label{fig:sync_delay}
\end{figure}

In Figure \ref{fig:sync_delay}, the synchronization delay is illustrated as a function of the overhead.  

There are two important results:  First, for analog BF, the omni-directional transmissions
(options ODx) outperform directional transmissions (Option DDx).  This is consistent
with the analysis in Section~\ref{sec:delAn}:  since the directional transmissions
were using the minimum $T_{\rm sig}$, the omni-directional transmissions provided a 
benefit (see the second part of the max in \eqref{eq:Dsbndana}).
Second, we see that digital reception (Option ODigDig) in both SNR regimes
significantly outperforms other schemes for all overhead ratios.
The digital BF curve accounts for the fact that we are using low-resolution
quantization to keep the power consumption roughly constant.
The gain with digital BF is due to its ability to ``look'' into multiple directions at once while exploiting directionality at both RX and TX,
and
is clearly observable when comparing the high SNR curves of DDO and ODigDig, as predicted by \eqref{eq:Dsbnddig}.

\subsection{Random Access Delay} \label{sec:raDelay}

\begin{figure}
\centering {\includegraphics[width=0.9\linewidth]{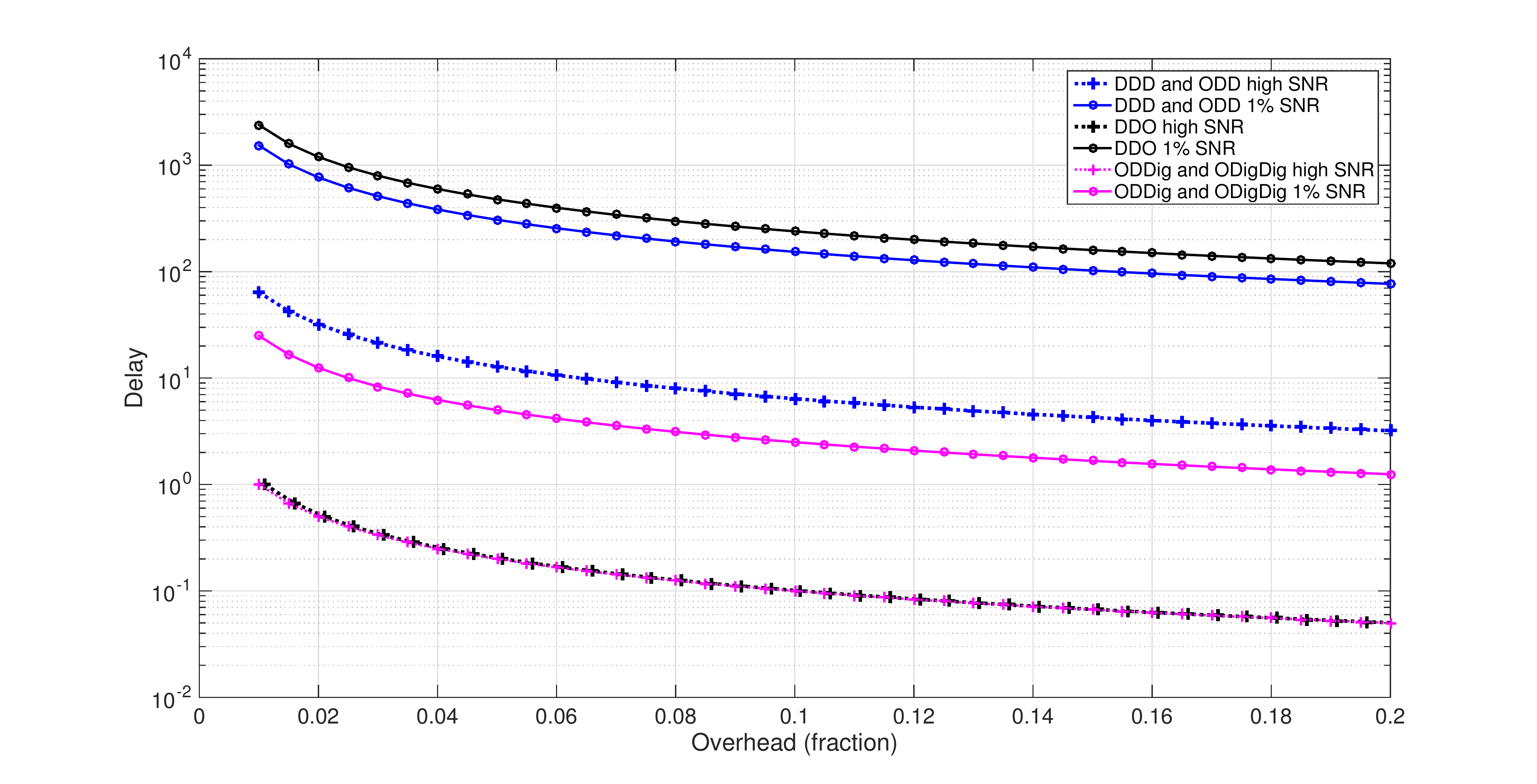}}
\caption{Synchronization delay in ms versus signal overhead for the minimum Sync signal duration $T_{\rm sig,RA}^{min} = 10$~${\rm \mu}$s.}
\label{fig:rach_delay}
\end{figure}

For the Random access phase, we perform a detection procedure similar to that of synchronization.
We consider the same SNR regimes for edge and high SNR users as before,
and we can compute the RA overhead as
\beq \label{eq:ovRAana}
    \phi_{\rm ov, \textsc RA} = T_{\rm sig,\textsc RA}/T_{\rm per,\textsc RA}.
\eeq
It should be noted that in digital BF or hybrid BF with more than one stream,  a lower overhead may be possible:  
With digital or multi-stream hybrid BF, in the same time slot of any 
RA period, a BS can theoretically schedule uplink transmissions 
from UEs that are already connected.
This is not possible with analog BF with one stream since the BS 
in that case can only ``look"  in one direction at a time.
Since the RA signals occupy a total frequency of 
$N_{\rm div}W_{\rm sig,\textsc RA}$, the total overhead with digital BF or multi-stream hybrid BF is

\beq \label{eq:ovRADig}
    \phi_{\rm ov, \textsc RA} = W_{\rm sig,\textsc RA}T_{\rm sig,\textsc RA}/
        (W_{\rm tot}T_{\rm per,\textsc RA}).
\eeq

Since this multiplexing would require a potentially larger dynamic range, possibly not afforded by low-resolution ADCs, the analysis below will forgo this potential improvement until further investigation can be conducted.
In the meantime, we will conservatively assume that the 
overhead in \eqref{eq:ovRAana} applies to both digital and analog BF.
The detection delay is then given by $
    D_{\rm RA} = KLT_{\rm per,RA} = \frac{KLT_{\rm sig,RA}}{\phi_{\rm ov,RA}}.$

Also, the same minimum duration $T_{\rm sig,RA}^{min} = 10~\mu$s is considered.
What is inherently different in RA, compared to the synchronization phase, is that the beam space is always smaller.
This is due to the fact that RA follows the synchronization phase.
Therefore, the UE has already obtained the correct angle to the BS and transmits its RA preamble only in that direction.
Also, the number of the delay hypotheses is considerably smaller and is a function of the propagation round trip time.
This is because we can assume that by detecting the synchronization signal, the UE learns where in the frame the BS expects the RA requests - see Appendix \ref{sec:simParam}.

Figure \ref{fig:rach_delay} and Table \ref{tab:delays} summarize the results of our RA simulations.
As in synchronization, to achieve both low overhead and delay, digital RX is always a better choice:
ODdig/ODigDig in both SNR regimes (red curves) outperform, in some cases  by orders of magnitude,
DDO (green curves) and DDD/ODD (blue curves).
Note that in all the schemes we calculated the overhead assuming that for the duration of $T_{\rm sig,RA}^{min}$
the BS expects to receive only RA requests and nothing else.
However, we need to stress that with digital BF and hence frequency multiplexing, the BS can receive both RA and data at the same time.
This reduces the overhead and increases even more the benefits of using digital BF.
Of course, high performance and low overhead come at the cost of large complexity and power consumption.
Similarly to the Sync phase, we again see a dramatic reduction of delay with digital BF at the BS.

\section{Findings and Conclusions} \label{concl}

A key challenge for initial access in mmWave cellular is the need for the BS and UE
to not only discover one another, but also determine the initial directions of communication.
We have analyzed various design options for this directional search in
both the synchronization phase where the UE discovers the BS and the random access
phase where the BS detects the access request from the UE.
Our analysis demonstrates several key findings:
\begin{itemize}

\item \emph{Cost of directionality:}  While directional transmissions are essential for mmWave,
they can significantly delay initial access.  
Our analysis in Section~\ref{sec:delAn} reveals that, with highly directional
beams at the BS, the synchronization delay grows linearly with the beamforming gain.  
With the large number of antenna elements projected to be employed in mmWave systems, 
this can enormously increase the delay.  This is confirmed in our simulations.

\item \emph{Omni-directional transmission of the synchronization signal:}
When the transmitted signal duration $T_{\rm sig,sync}$ is unbounded, there is theoretically no difference in delay whether the BS transmits the
synchronization signals directionally or omni-directionally.
When the BS uses directional transmissions,
the synchronization signal should be transmitted very frequently with each transmission occupying a very short duration.
However, our evaluations suggest that the necessary
Sync signal duration for directional transmissions to perform as well as omni-directional ones may be too short for practical implementations.
Therefore, when using a signal duration greater than or equal to a lower bound $T_{\rm sig, sync}^{min}$, omni-directional transmission
of the synchronization signal may have significant benefits.

\item \emph{Value of digital BF}:  Fully digital architectures
can dramatically improve the delay even further.  Since the receiver
can ``look" in all directions at once, the delay from sequential scanning is removed.  
In addition, in the random access phase, the BS can use digital BF to multiplex
other channels at the same time (but on different frequencies) to reduce the overhead.
The added power consumption
of fully digital transceivers can be compensated by using low-resolution quantizers
at minimal performance cost.

\end{itemize}

With digital BF we can obtain access delays of a few milliseconds -- an order
of magnitude faster than the control plane latency target in LTE of 50~ms \cite{3GPP36913}.
This possibility suggests that mmWave systems enabled with low-resolution fully digital transceivers
can not only offer dramatically improved data rates and lower data plane latency but significantly
reduced control plane latency as well.  Future work can study how systems can be best designed
to exploit this capability, particularly for inter-RAT handover, aggressive use of idle modes
and fast recovery from link failure.

\appendices
\section{Derivation of the GLRT} \label{sec:GLRTderiv}

Under the signal model in \eqref{eq:ruvalpha}, $\rbf_{k\ell d}$ is conditionally Gaussian given
the parameters $\psi_{k d}$, $\ell_0$ and $\tau_{k\ell d}$.
Therefore, the negative log likelihood is
\beqa
    \lefteqn{ -\ln p(\rbf_{k\ell d}|\tau_{k\ell d},\psi_{k d},\ell_0) } \nonumber \\
    && =
        \frac{1}{\tau_{k\ell d}} \|\rbf_{k\ell d} - \psi_{k d}\delta_{\ell,\ell_0}\sbf_{\ell d}\|^2
        + M\ln(2\pi\tau_{k\ell d}).
\eeqa
Now define the minimum log likelihoods
\begin{subequations} \label{eq:Lamkell}
\begin{flalign}
&&\Lambda^{k\ell d}_0(\ell_0) & := \min_{\tau_{k\ell d},\psi_{k d} = 0}  -\ln p(\rbf_{k\ell d}|\tau_{k\ell d},\psi_{kd},\ell_0)&&\\
&&\Lambda^{k\ell d}_1(\ell_0) & := \min_{\tau_{k\ell d},\psi_{k d}}
     -\ln p(\rbf_{k\ell d}|\tau_{k\ell d},\psi_{k d},\ell_0) &&
\end{flalign}
\end{subequations}
For $\Lambda^{k\ell d}_0(\ell_0)$, the minimum over $\tau_{k\ell}$ occurs at
$
    \tau_{k\ell d} = \|\rbf_{k\ell d}\|^2,
$
and we obtain
\beq \label{eq:lam0}
    \Lambda^{k\ell d}_0(\ell_0) = M\ln\left(2\pi e \|\rbf_{k\ell d}\|^2 \right).
\eeq
When $\ell\neq \ell_0$, we obtain the same expression for
$\Lambda^{k\ell d}_1(\ell_0)$,
\beq \label{eq:lam1a}
    \Lambda^{k\ell d}_1(\ell_0) = M\ln\left(2\pi e \|\rbf_{k\ell d}\|^2 \right), \quad
    \ell \neq \ell_0.
\eeq
For $\ell=\ell_0$, we first minimize over $\psi_{kd}$ and obtain
$
    \psi_{k d} = \frac{\sbf_{\ell d}^*\rbf_{k\ell d}}{\|\sbf_{\ell d}\|^2}.
$

Next, we perform the minimization over $\tau_{k\ell d}$ similar to the $\Lambda^{k\ell d}_0$ case which results in

$
    \tau_{k\ell d} = \|\rbf_{k\ell d}\|^2 - \frac{|\sbf_{\ell d}^*\rbf_{k\ell d}|^2}{\|\sbf_{\ell d}\|^2},
$
and finally we obtain

\beq \label{eq:lam1b}
    \Lambda^{k\ell d}_1(\ell_0) = M\ln\left(2\pi e \left[\|\rbf_{k\ell d}\|^2
        - \frac{|\sbf_{\ell d}^*\rbf_{k\ell d}|^2}{\|\sbf_{\ell d}\|^2} \right]\right).
\eeq
Therefore, if we define
\beq \label{eq:Tkell}
    T^{k\ell d}(\ell_0) := \Lambda^{k\ell d}_1(\ell_0) - \Lambda^{k\ell d}_0(\ell_0),
\eeq
we get
\beq \label{eq:Tkellln}
    T^{k\ell d}(\ell_0) = M \ln \left(1 - \rho_{k\ell d}\right)\delta_{\ell,\ell_0},
\eeq
where $\rho_{k\ell}$ is the correlation in \eqref{eq:rhokl}.

Under the assumption that the noise vectors are independent on different
transmissions,
\beq \label{eq:logpsum}
     \ln p(\rbf|\taubf,\psibf,\ell_0) =
     \sum_{k=1}^K \sum_{\ell=1}^L \sum_{d=1}^{N_{\rm div}} \ln p(\rbf_{k\ell d}|\tau_{k\ell d},\psi_{k d},\ell_0).
\eeq
Due to \eqref{eq:logpsum}, the negative log likelihoods \eqref{eq:MLhyp}
are given by
\[
    \Lambda_i = \min_{\ell_0}  \sum_{k=1}^K \sum_{\ell=1}^L \sum_{d=1}^{N_{\rm div}} \Lambda^{k\ell d}_i(\ell_0).
\]
Therefore,
we can apply \eqref{eq:Tkell} and \eqref{eq:Tkellln} to obtain
\beqa
    \lefteqn{\Lambda_1 - \Lambda_0 = \min_{\ell_0} \sum_{k=1}^K \sum_{\ell=1}^L  \sum_{d=1}^{N_{\rm div}} T^{k\ell d}(\ell_0)} \\
    &=& \min_{\ell_0}M\sum_{k=1}^K \sum_{\ell=1}^L \sum_{d=1}^{N_{\rm div}} \ln\left(1-\rho_{k\ell d}\right)\delta_{\ell,\ell_0}.
    \label{eq:Tsum}
\eeqa
It can then be seen that the minimization of \eqref{eq:Tsum} occurs at $\ellhat_0$
given in \eqref{eq:ellhat} with minimum value \eqref{eq:Trho}.

\section{Simulation Parameter Selection Details}\label{sec:simParam}
Here we provide more details on the
selection of the simulation parameters.
We also hint at some of the considerations that should be taken into account in selecting parameter values for practical systems.

\paragraph{Signal parameters}
We assume that both the synchronization
signal and the random access signal are divided into
$N_{\rm div}$ narrow-band subsignals sent over different
frequency bands to provide frequency diversity.
Our experiments indicated that $N_{\rm div}=4$ provides a good
tradeoff between frequency diversity and coherent combining.

For the length of the signals, we tried three different cases: a very short signal of duration  $T_{\rm sig}=$ 10 $\mu$s
and two longer ones with $T_{\rm sig}=$ 50 $\mu$s and $T_{\rm sig}=$ 100 $\mu$s.
All of them are sufficiently small for the channel to be coherent even at the
very high frequencies of mmWave communication.
For instance, if we take a moderately mobile UE with a velocity of $30$ km/h ($\sim 18.6$ mph), at 28 GHz
the maximum Doppler shift of the mmWave channel is $\approx 780$ Hz.
Therefore, the coherence time is $\gg 100~\mu$s, which is the maximum of the signal lengths we consider.
Furthermore, typical delay spreads in a mmWave outdoor setting within a narrow angular region are $< 30\; \text{ns}$ \cite{Rappaport:12-28G, McCartRapICC15}.
This means that if we take a subsignal bandwidth of $W_{\rm sig}$ = 1 MHz, the channel will be relatively flat across this band.

Since the transmission occurs every $T_{\rm per}$ seconds,
it has to account for the ability to track UEs in motion and resynchronize UEs for which the signal is lost due to sudden
changes in the direction.
Hence, $T_{\rm per}$ needs to be frequent enough to ensure fast synchronization and resynchronization times.
However, depending on the length of the signal, a very frequent signal transmission may
introduce high overhead.
Therefore, there is a tradeoff between overhead and delay in choosing the frequency of signal transmission.
The overhead is defined as the ratio of the signal length $T_{\rm sig}$ over the transmission period $T_{\rm per}$.
Therefore, if we fix $T_{\rm per}$, we will get different overhead values for different signal lengths.
For instance, for $T_{\rm per} = 1$~ms we get $1 \%$, $ 5 \%$ and $10 \%$ overhead for
$T_{\rm sig} =$ 10, 50 and 100 $\mu$s, respectively.
In Figures \ref{fig:sync_delay} and \ref{fig:rach_delay}, we show the delay as a function of the overhead
which can be interpreted as varying $T_{\rm per}$ from very sparse to very frequent.
As mentioned in Section \ref{sec:simSNR} the path loss based SNR is calculated using the pre
beamforming data SNR, i.e., the SNR at which
the UE will communicate data packets with the BS using all the available bandwidth,
without considering the directional antenna gains.
This is different from the narrow band SNR of the synchronization or the RA signals and is equal to
$\gamma_0 = \frac{P_{\rm omni}}{N_0 W_{\rm tot}}$, where
$P_{\rm omni}$ is the omnidirectional received power, and  $N_0$ is the noise power spectral density.
Based on this, the pre-beamforming SNR of the synchronization signal can be derived as:
$
    \gamma_{sync} = \gamma_0T_{\rm sig}W_{\rm tot}.
$
Similarly, if we consider the effect of beamforming, the signal SNR can be computed as
$
    \gamma_{\rm sync,dir} = \gamma_0T_{\rm sig}W_{\rm tot}G_{\rm tx}^{\rm sync}G_{\rm rx},
$
where $G_{\rm tx}^{\rm sync}$ and $G_{\rm rx}$ are the antenna gains of the transmitter and receiver, respectively.

\paragraph{False alarm target}
In this simulation, we will assume that the total false alarm rate is at most $R_{\textsc FA}$ = 0.01 false alarms per transmission period.
Thus, the false alarm rate per delay hypothesis
must be $P_{\textsc FA} \leq R_{\textsc FA}/N_{\rm hyp}$ where $N_{\rm hyp}$ is the number
of hypotheses that will be tested per transmission period.  The number of hypotheses are given by the product
$
    N_{\rm hyp} = N_{\rm sig}N_{\rm dly}N_{\textsc FO},
$
where $N_{\rm dly}$ is the number of delay hypotheses per transmission period $T_{\rm per}$,
$N_{\rm sig}$ is the number of possible signals that can be transmitted by the
BS ($N_{\rm sync}$ in the synchronization step)
or the UE ($N_{\rm rach}$ in the random access step),
and $N_{\textsc FO}$ is the number of frequency offsets.

To calculate $N_{\rm dly}$, we have to distinguish between initial synchronization and random access.
In the Sync phase, the UE must search over delays in the range $\tau \in [0,T_{\rm per}]$.
Assuming the correlations are computed
at twice the bandwidth, we will have
$N_{\rm dly} = 2W_{\rm sig} T_{\rm per} = 2 (10)^6 \times 5 (10)^{-3} = 10^4$,
if we take an overhead resulting into a $T_{\rm per} = 5$ ms.
Note that  $5$ ms is the Sync transmission period used in the 3GPP LTE systems \cite{3GPP36.300,3GPP36.321,3GPP36.331}.
After the synchronization signal is received at the UE, the UE must trigger the random access
by transmitting a random preamble back to the BS.
Thus, the BS will receive the preamble after one
full round trip time after the signal transmission.
In order to calculate the round trip time, we assume a cell radius of 100 m.
As a result the round trip time is found to be $T_{D}=2\frac{100}{3\times10^8} = 0.66~\mu s$ and
we have $N_{\rm dly} = 2W_{\rm sig} T_{D} = 2 (10)^6 \times 0.66 (10)^{-6} = 1.33$.

For the number of synchronization signals, we will take $N_{\rm sig} = N_{\rm sync} = 3$,
which is the same as the number of primary synchronization signals in the current LTE system.
However, we may need to increase this number to accommodate more cell IDs in a dense BS deployment.
For the random access signal, we use $N_{\rm sig} = N_{\rm rach}=64$ which is also used in the current LTE system.
The higher the number of preambles, the higher the detection resolution and the lower the likelihood
of contention during random access.
To estimate the number of frequency offsets, suppose that the initial
frequency offset can be as much as 1 ppm at 28~GHz.
This will result in a frequency offset of $28(10)^3$ Hz.
The maximum Doppler shift will add approximately 780~Hz, giving a total
initial frequency offset error of $\Delta f_{\rm max}=28 + 0.78$~kHz.
For the channel not to rotate
more than 90$^\circ$ over the period of maximum $T_{\rm sig}=100$ $\mu$s, we need
a frequency accuracy of $\Delta f= 10^{4}/4$.  Since $2\Delta f_{max}/\Delta f = 23.0$,
it will suffice to take $N_{\textsc FO}=23$ frequency offset hypotheses.  Hence, the target
FA rate for initial access should be
\[
    P_{\textsc FA,DL} = \frac{R_{\textsc FA}}{N_{\rm hyp}} = \frac{0.01}{10^4 (3) 23} = 1.4493(10)^{-8},
\]
and the FA rate for random access is:

\[
    P_{\textsc FA,UL} = \frac{R_{\textsc FA}}{N_{\rm hyp}} = \frac{0.01}{1.33 (64) 23} =   5.1079(10)^{-6}.
\]
\paragraph{Antenna pattern} We assume a set of two dimensional antenna arrays at both the BS and the UE.
On the BS and the UE sides, the arrays consist of $8 \times 8$, and $4 \times 4$ elements, respectively.
The spacing of the elements is set at half the signal wavelength.
These antenna patterns were used in \cite{AkdenizCapacity:14} showing excellent system capacity for small cell urban deployments.

\bibliographystyle{IEEEtran}
\bibliography{bibl}

\newcommand{\SortNoop}[1]{}
\begin{thebibliography}{10}
\providecommand{\url}[1]{#1}
\csname url@samestyle\endcsname
\providecommand{\newblock}{\relax}
\providecommand{\bibinfo}[2]{#2}
\providecommand{\BIBentrySTDinterwordspacing}{\spaceskip=0pt\relax}
\providecommand{\BIBentryALTinterwordstretchfactor}{4}
\providecommand{\BIBentryALTinterwordspacing}{\spaceskip=\fontdimen2\font plus
\BIBentryALTinterwordstretchfactor\fontdimen3\font minus
  \fontdimen4\font\relax}
\providecommand{\BIBforeignlanguage}[2]{{%
\expandafter\ifx\csname l@#1\endcsname\relax
\typeout{** WARNING: IEEEtran.bst: No hyphenation pattern has been}%
\typeout{** loaded for the language `#1'. Using the pattern for}%
\typeout{** the default language instead.}%
\else
\language=\csname l@#1\endcsname
\fi
#2}}
\providecommand{\BIBdecl}{\relax}
\BIBdecl

\bibitem{BarHosAsilomar:15}
C.~N. Barati, S.~A. Hosseini, M.~Mezzavilla, P.~Amiri-Eliasi, S.~Rangan,
  T.~Korakis, S.~S. Panwar, and M.~Zorzi, ``Directional initial access for
  millimeter wave cellular systems,'' in \emph{Proc. of Asilomar Conf. on
  Signals, Syst. \& Computers}, Nov. 2015, pp. 307--311.

\bibitem{KhanPi:11-CommMag}
F.~Khan and Z.~Pi, ``{An introduction to millimeter-wave mobile broadband
  systems},'' \emph{IEEE Commun. Mag.}, vol.~49, no.~6, pp. 101 -- 107, June
  2011.

\bibitem{rappaportmillimeter}
T.~S. Rappaport, S.~Sun, R.~Mayzus, H.~Zhao, Y.~Azar, K.~Wang, G.~N. Wong,
  J.~K. Schulz, M.~Samimi, and F.~Gutierrez, ``{Millimeter Wave Mobile
  Communications for 5G Cellular: It Will Work!}'' \emph{IEEE Access}, vol.~1,
  pp. 335--349, May 2013.

\bibitem{RanRapE:14}
S.~Rangan, T.~S. Rappaport, and E.~Erkip, ``Millimeter-wave cellular wireless
  networks: Potentials and challenges,'' \emph{Proc. IEEE}, vol. 102, no.~3,
  pp. 366--385, Mar. 2014.

\bibitem{andrews2014will}
J.~Andrews, S.~Buzzi, W.~Choi, S.~Hanly, A.~Lozano, A.~Soong, and J.~Zhang,
  ``What will 5{G} be?'' \emph{IEEE J. Sel. Areas Commun.}, vol.~32, no.~6, pp.
  1065--1082, June 2014.

\bibitem{ghosh2014millimeter}
A.~Ghosh, T.~A. Thomas, M.~C. Cudak, R.~Ratasuk, P.~Moorut, F.~W. Vook, T.~S.
  Rappaport, G.~MacCartney, S.~Sun, and S.~Nie, ``Millimeter wave enhanced
  local area systems: A high data rate approach for future wireless networks,''
  \emph{IEEE J. Sel. Areas Commun.}, vol.~32, no.~6, pp. 1152--1163, June 2014.

\bibitem{DehosG:14}
C.~Dehos, J.~L. Gonzalez, A.~D. Domenico, D.~Ktenas, and L.~Dussopt,
  ``Millimeter-wave access and backhauling: the solution to the exponential
  data traffic increase in 5{G} mobile communications systems?'' \emph{IEEE
  Commun. Mag.}, vol.~52, no.~9, pp. 88--95, Sept. 2014.

\bibitem{AkdenizCapacity:14}
M.~Akdeniz, Y.~Liu, M.~Samimi, S.~Sun, S.~Rangan, T.~Rappaport, and E.~Erkip,
  ``Millimeter wave channel modeling and cellular capacity evaluation,''
  \emph{IEEE J. Sel. Areas Commun.}, vol.~32, no.~6, pp. 1164--1179, June 2014.

\bibitem{BaiHeath:15}
T.~Bai and R.~Heath, ``Coverage and rate analysis for millimeter-wave cellular
  networks,'' \emph{IEEE Trans. Wireless Commun.}, vol.~14, no.~2, pp.
  1100--1114, Feb. 2015.

\bibitem{shokri2015millimeter}
H.~Shokri-Ghadikolaei, C.~Fischione, G.~Fodor, P.~Popovski, and M.~Zorzi,
  ``Millimeter wave cellular networks: A {MAC} layer perspective,'' \emph{IEEE
  Trans. Commun.}, vol.~63, no.~10, pp. 3437--3458, Oct. 2015.

\bibitem{giordani:16multi}
M.~Giordani, M.~Mezzavilla, S.~Rangan, and M.~Zorzi, ``Multi-connectivity in
  5{G} mmwave cellular networks,'' \emph{submitted to IEEE Commun. Magazine
  (available at arxiv.org/abs/1605.00105)}, 2016.

\bibitem{BocHLMP:14}
F.~Boccardi, R.~Heath, A.~Lozano, T.~Marzetta, and P.~Popovski, ``Five
  disruptive technology directions for 5{G},'' \emph{IEEE Commun. Mag.},
  vol.~52, no.~2, pp. 74--80, Feb. 2014.

\bibitem{levanen2014radio}
T.~Levanen, J.~Pirskanen, and M.~Valkama, ``Radio interface design for
  ultra-low latency millimeter-wave communications in {5G} era,'' in
  \emph{Proc. IEEE Globecom Workshops}, Dec. 2014, pp. 1420--1426.

\bibitem{BarHosCellSearch:TWC15}
C.~Barati, S.~Hosseini, S.~Rangan, P.~Liu, T.~Korakis, S.~Panwar, and
  T.~Rappaport, ``Directional cell discovery in millimeter wave cellular
  networks,'' \emph{IEEE Trans. Wireless Commun.}, vol.~14, no.~12, pp.
  6664--6678, Dec. 2015.

\bibitem{KhanPi:11}
F.~Khan and Z.~Pi, ``{Millimeter-wave {M}obile {B}roadband ({MMB}):
  {U}nleashing 3-300GHz Spectrum},'' in \emph{Proc. IEEE Sarnoff Symposium},
  May 2011.

\bibitem{Abbas:16BF}
W.~B. Abbas and M.~Zorzi, ``Towards an appropriate receiver beamforming scheme
  for millimeter wave communication: A power consumption based comparison,'' in
  \emph{Proc. European Wireless conference 2016 (EW16)}, May 2016.

\bibitem{KohReb:07}
K.-J. Koh and G.~Rebeiz, ``{0.13-${\mu}$m CMOS Phase Shifters for X-, Ku-, and
  K-Band Phased Arrays},'' \emph{IEEE J. Solid-State Circuts}, vol.~42, no.~11,
  pp. 2535--2546, Nov. 2007.

\bibitem{KohReb:09}
K.-J. Koh, J.~May, and G.~Rebeiz, ``A millimeter-wave (40-45 {GH}z) 16-element
  phased-array transmitter in 0.18-$\mu$m {S}i{G}e {B}i{CMOS} technology,''
  \emph{IEEE J. Solid-State Circuts}, vol.~44, no.~5, pp. 1498--1509, May 2009.

\bibitem{GuanHaHa:04}
X.~Guan, H.~Hashemi, and A.~Hajimiri, ``{A fully integrated 24-GHz
  eight-element phased-array receiver in silicon},'' \emph{IEEE J. Solid-State
  Circuts}, vol.~39, no.~12, pp. 2311--2320, Dec. 2004.

\bibitem{Heath:partialBF}
A.~Alkhateeb, O.~E. Ayach, G.~Leus, and J.~Robert W.~Heath, ``Hybrid precoding
  for millimeter wave cellular systems with partial channel knowledge,'' in
  \emph{Proc. Inform. Theory and Applicat. Workshop (ITA)}, Feb. 2013.

\bibitem{SunRap:cm14}
S.~Sun, T.~Rappaport, R.~Heath, A.~Nix, and S.~Rangan, ``{MIMO} for
  millimeter-wave wireless communications: beamforming, spatial multiplexing,
  or both?'' \emph{IEEE Commun. Mag.}, vol.~52, no.~12, pp. 110--121, Dec.
  2014.

\bibitem{Madhow:ADC}
H.~Zhang, S.~Venkateswaran, and U.~Madhow, ``Analog multitone with interference
  suppression: Relieving the {ADC} bottleneck for wideband 60 {GHz} systems,''
  in \emph{Proc. IEEE Global Commun. Conference (GLOBECOM)}, Nov. 2012, pp.
  2305--2310.

\bibitem{Madhow:largeArray}
D.~Ramasamy, S.~Venkateswaran, and U.~Madhow, ``Compressive tracking with
  1000-element arrays: A framework for multi-{G}bps mm wave cellular
  downlinks,'' in \emph{Proc. Ann. Allerton Conf. on Commun., Control and
  Comp.}, Oct. 2012, pp. 690--697.

\bibitem{hassan2010analog}
K.~Hassan, T.~Rappaport, and J.~Andrews, ``Analog equalization for low power 60
  {GH}z receivers in realistic multipath channels,'' in \emph{Global
  Telecommunications Conference (GLOBECOM 2010), 2010 IEEE}, Dec. 2010.

\bibitem{Nitsche:15}
T.~Nitsche, A.~B. Flores, E.~W. Knightly, and J.~Widmer, ``Steering with eyes
  closed: mm-wave beam steering without in-band measurement,'' in \emph{Proc.
  IEEE Conference on Computer Communications (INFOCOM)}, Apr. 2015, pp.
  2416--2424.

\bibitem{capone:15context}
A.~Capone, I.~Filippini, and V.~Sciancalepore, ``Context information for fast
  cell discovery in mm-wave 5{G} networks,'' in \emph{Proc. 21th European
  Wireless Conference}, May 2015.

\bibitem{jung:15cellDtct}
H.~Jung, Q.~Li, and P.~Zong, ``Cell detection in high frequency band small cell
  networks,'' in \emph{Proc. of Asilomar Conf. on Signals, Syst. \& Computers},
  Nov. 2015, pp. 1762--1766.

\bibitem{Abbas:16context}
W.~B. Abbas and M.~Zorzi, ``Context information based initial cell search for
  millimeter wave 5{G} cellular networks,'' in \emph{Proc. European Conference
  on Networks and Communications (EuCNC 2016)}, Jun. 2016.

\bibitem{3GPP36.300}
3GPP, ``{Evolved Universal Terrestrial Radio Access (E-UTRA) and Evolved
  Universal Terrestrial Radio Access Network (E-UTRAN); Overall description;
  Stage 2},'' TS 36.300 (release 10), 2010.

\bibitem{3GPP36.321}
------, ``{Evolved Universal Terrestrial Radio Access (E-UTRA) and Evolved
  Universal Terrestrial Radio Access Network (E-UTRAN); Medium Access Control
  (MAC) protocol specification},'' TS 36.321, 2015.

\bibitem{3GPP36.331}
------, ``{Evolved Universal Terrestrial Radio Access (E-UTRA) and Evolved
  Universal Terrestrial Radio Access Network (E-UTRAN); Radio Resource Control
  (RRC) protocol specification},'' TS 36.331, 2015.

\bibitem{Dahlman:07}
E.~Dahlman, S.~Parkvall, J.~Sk{\"o}ld, and P.~Beming, \emph{{3G} Evolution:
  {HSPA} and {LTE} for Mobile Broadband}.\hskip 1em plus 0.5em minus
  0.4em\relax Oxford, UK: Academic Press, 2007.

\bibitem{Lozano:07}
A.~Lozano, ``Long-term transmit beamforming for wireless multicasting,'' in
  \emph{Proc. IEEE Int. Conference on Acoust., Speech, and Signal Process.},
  vol.~3, Apr. 2007, pp. III--417--III--420.

\bibitem{jeong2015random}
C.~Jeong, J.~Park, and H.~Yu, ``Random access in millimeter-wave beamforming
  cellular networks: issues and approaches,'' \emph{IEEE Commun. Mag.},
  vol.~53, no.~1, pp. 180--185, Jan. 2015.

\bibitem{VanTrees:01a}
H.~L. {Van Trees}, \emph{Detection, Estimation and Modulation Theory, Part
  I}.\hskip 1em plus 0.5em minus 0.4em\relax New York, NY: Wiley, 2001.

\bibitem{shokri2015beamsearch}
H.~Shokri-Ghadikolaei, L.~Gkatzikis, and C.~Fischione, ``Beam-searching and
  transmission scheduling in millimeter wave communications,'' in \emph{Proc.
  IEEE International Conference on Communications (ICC)}, June 2015, pp.
  1292--1297.

\bibitem{bogale2015hybridADestimation}
T.~E. Bogale, L.~B. Le, and X.~Wang, ``Hybrid analog-digital channel estimation
  and beamforming: Training-throughput tradeoff,'' \emph{IEEE Trans. Commun.},
  vol.~63, no.~12, pp. 5235--5249, Dec. 2015.

\bibitem{Rappaport:12-28G}
Y.~Azar, G.~Wong, K.~Wang, R.~Mayzus, J.~Schulz, H.~Zhao, F.~Gutierrez,
  D.~Hwang, and T.~Rappaport, ``28 {GH}z propagation measurements for outdoor
  cellular communications using steerable beam antennas in {N}ew {Y}ork
  {C}ity,'' in \emph{Proc. IEEE International Conference on Communications
  (ICC)}, June 2013, pp. 5143--5147.

\bibitem{McCartRapICC15}
G.~Maccartney, M.~Samimi, and T.~Rappaport, ``Exploiting directionality for
  millimeter-wave wireless system improvement,'' in \emph{Proc. IEEE
  International Conference on Communications (ICC)}, June 2015, pp. 2416--2422.

\bibitem{Rappaport:28NYCPenetrationLoss}
H.~Zhao, R.~Mayzus, S.~Sun, M.~Samimi, J.~Schulz, Y.~Azar, K.~Wang, G.~Wong,
  F.~Gutierrez, and T.~Rappaport, ``28 {G}hz millimeter wave cellular
  communication measurements for reflection and penetration loss in and around
  buildings in {N}ew {Y}ork {C}ity,'' in \emph{Proc. IEEE International
  Conference on Communications (ICC)}, June 2013, pp. 5163--5167.

\bibitem{Samimi:AoAD}
M.~Samimi, K.~Wang, Y.~Azar, G.~N. Wong, R.~Mayzus, H.~Zhao, J.~K. Schulz,
  S.~Sun, F.~Gutierrez, and T.~S. Rappaport, ``28 {GHz} angle of arrival and
  angle of departure analysis for outdoor cellular communications using
  steerable beam antennas in {N}ew {Y}ork {C}ity,'' in \emph{Proc. IEEE Veh.
  Technology Conference (VTC)}, June 2013.

\bibitem{mo2014high}
J.~Mo and R.~W. Heath, ``High {SNR} capacity of millimeter wave {MIMO} systems
  with one-bit quantization,'' in \emph{Proc. Inform. Theory and Applicat.
  Workshop (ITA)}, Feb. 2014.

\bibitem{mo2014channel}
J.~Mo, P.~Schniter, N.~G. Prelcic, and R.~W. Heath~Jr, ``Channel estimation in
  millimeter wave {MIMO} systems with one-bit quantization,'' in \emph{Proc. of
  Asilomar Conf. on Signals, Syst. \& Computers}, Nov. 2014, pp. 957--961.

\bibitem{OrhanITA15}
O.~Orhan, E.~Erkip, and S.~Rangan, ``Low power analog-to-digital conversion in
  millimeter wave systems: Impact of resolution and bandwidth on performance,''
  in \emph{Proc. Inform. Theory and Applicat. Workshop (ITA)}, Feb. 2015, pp.
  191--198.

\bibitem{razavi2001design}
B.~Razavi, \emph{Design of Analog {CMOS} Integrated Circuits}.\hskip 1em plus
  0.5em minus 0.4em\relax McGraw Hill, 2003.

\bibitem{GershoG:92}
A.~Gersho and R.~M. Gray, \emph{Vector Quantization and Signal
  Compression}.\hskip 1em plus 0.5em minus 0.4em\relax Boston, MA: Kluwer Acad.
  Pub., 1992.

\bibitem{FletcherRGR:07}
A.~K. Fletcher, S.~Rangan, V.~K. Goyal, and K.~Ramchandran, ``Robust predictive
  quantization: Analysis and design via convex optimization,'' \emph{IEEE J.
  Sel. Topics Signal Process.}, vol.~1, no.~4, pp. 618--632, Dec. 2007.

\bibitem{3GPP36913}
3GPP, ``{Requirements for further advancements for Evolved Universal
  Terrestrial Radio Access (E-UTRA) (LTE-Advanced)},'' TR 36.913 (release 11),
  2012.

\end{thebibliography}

\end{document}